%% file: evolved-thresholds.tex
\documentclass[11pt]{article}
\usepackage{arxiv}

\usepackage{graphicx}
\usepackage{xcolor}
\usepackage{hyperref}
\usepackage{setspace}

\usepackage{booktabs} 

\newcommand{\Comments}{1}
\newcommand{\mynote}[2]{\ifnum\Comments=1\textcolor{#1}{#2}\fi}
\newcommand{\mytodo}[2]{\ifnum\Comments=1%
  \todo[linecolor=#1!80!black,backgroundcolor=#1,bordercolor=#1!80!black]{#2}\fi}

\newcommand{\simpath}[1]{\texttt{#1}}

\newtheorem{goal}{Domain Goal}

\renewcommand{\shorttitle}{Analysis of Evolved Response Thresholds}

\begin{document}

\date{}

\title{Analysis of Evolved Response Thresholds for Decentralized Dynamic Task Allocation} 

\author{\href{https://orcid.org/0000-0001-8521-6705}{\includegraphics[scale=0.06]{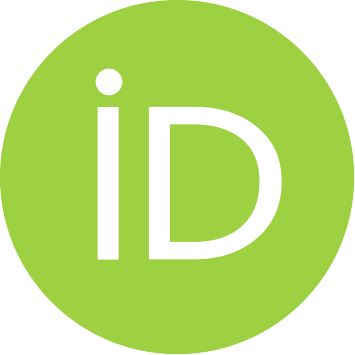}\hspace{1mm}H. David Mathias}\\
University of Wisconsin - La Crosse \\
Department of Computer Science\\
La Crosse, Wisconsin, USA\\
\texttt{dmathias@uwlax.edu} \\
\And
Annie S. Wu \\
University of Central Florida \\
Department of Computer Science\\
Orlando, Florida, USA \\
\texttt{aswu@cs.ucf.edu} \\
\And
Daniel Dang \\
Whitman College \\
Department of Computer Science\\
Walla Walla, Washington, USA \\
\texttt{dangd@whitman.edu} \\
}

\newcommand{\shortauthors}{Mathias, Wu, \& Dang}

\maketitle

\begin{abstract}
We investigate the application of a multi-objective genetic algorithm to the 
problem of task allocation in a self-organizing, decentralized, threshold-based swarm.  
Each agent in our system is capable of performing four tasks with a response 
threshold for each, and we seek to assign response threshold values to all of 
the agents a swarm such that the collective behavior of the swarm is optimized.
Random assignment of threshold values
according to a uniform distribution is known to be effective;
however, this method does not consider features of particular problem instances.
Dynamic response thresholds have some flexibility to address problem specific 
features through real-time adaptivity, often improving swarm performance.

In this work, we use a multi-objective genetic algorithm to evolve response thresholds
for a simulated swarm engaged in a dynamic task allocation problem: two-dimensional collective tracking.  
We show that evolved thresholds not only outperform
uniformly distributed thresholds and dynamic thresholds but achieve nearly optimal performance 
on a variety of tracking problem instances (target paths).
More importantly, we demonstrate that thresholds evolved for one of several problem instances generalize to
all other problem instances eliminating the need to evolve new thresholds for each problem to be solved.
We analyze the properties that allow these paths to serve as universal training instances and show that they are 
quite natural.
\end{abstract}


\onehalfspace

\input{sec-intro}

\input{sec-bg}

\input{sec-model}

\input{sec-ga}


\input{sec-results}

\input{sec-conclusion}

\section*{Acknowledgements}
This work was supported by the National Science Foundation 
under Grant No. IIS1816777.

\newpage

\singlespace

\bibliographystyle{plain}
\bibliography{evolved-thresh} 

\end{document}

%% file: sec-intro.tex
\section{Introduction}
\label{sec:intro}

%
%
%

In this work, we use a multi-objective genetic algorithm (GA) to evolve agent 
response thresholds for a decentralized, threshold-based swarm. 
The decentralized and redundant qualities of swarms make them robust and 
scalable, but also make the coordination of the agents that make up a swarm 
a challenging problem.
The response threshold approach is modeled after
the division of labor in biological swarms and
is a commonly used approach for coordinating artificial swarms.
Optimal performance of such systems must address multiple goals and depends 
on effective assignment of threshold values among the agents of the swarm.
Because agents work collectively,
threshold assignments cannot be optimized locally
and the problem to be solved entails finding a collection of threshold
values that together generate optimal swarm performance for a given
set of task demands.
We show that a multi-objective GA (MOGA) is able to find near optimal
solutions to this multi-objective, large scale combinatoric problem and, more significantly,
we demonstrate that response thresholds evolved for some problem instances generalize
to provide near optimal performance for all other problem instances tested despite significant
differences in task demands between problem instances.

Eusocial insects, such as ants, bees, and wasps, are species with complex social structures within which individuals divide labor among various tasks such as foraging for food, foraging for nest building materials, and brood care.  Division of labor in the absence of centralized control of individual behaviors is a difficult problem that the eusocial insects have solved quite effectively \cite{jeanne1986}.  Response thresholds are one determinant of when an individual undertakes a task.  In the response threshold model, individuals sense environmental stimuli such as the amount of stored food or the temperature in the hive and act if the stimulus exceeds a threshold value.
Inter-individual variation, differences in when and how individuals respond to task demands, is an important mechanism for effective division of labor \cite{jeanson2014,weidenmuller2004,weidenmuller2019}.  In particular, variability in response thresholds serves to desynchronize activations by individuals, preventing the swarm from responding in lockstep.

Such natural swarm behaviors have inspired work in robotics for at least 30 years \cite{beni1992,theraulaz1991}.  An artificial
swarm consists of a number of agents working to achieve a common goal, usually through repetition of some number
of tasks.  Within this broad definition, there are many models defined by parameters such as inter-agent communication
and the mechanism used to achieve division of labor.  In a decentralized swarm, division of labor is achieved through
individual agent decisions made without a shared controller.  
In this work, we focus on swarms consisting of threshold based agents
and we assume no inter-agent communication.  

As in natural swarms, homogeneity in agent behaviors in artificial swarms
will result in little division of labor and poor goal attainment. 
Two main approaches have been used in artificial threshold-based swarms to 
promote heterogeniety in agent behaviors.
First,
threshold based systems can be probabilistic 
\cite{bonabeau1996,kalra2006,price2004};
agents act with some probability when their threshold is met.
The probabilistic action generates diversity in agent behaviors even if all
agents have the same threshold.
Second, and the approach we study here,
threshold based systems can be deterministic but variable;
behavioral variability is due to inter-agent variation in threshold values.
In this approach, it stands to reason that 
the method used to determine and assign 
response thresholds to agents could significantly impact the success of the 
swarm.  
Surprisingly little work, however, has examined this important question 
\cite{campbell2011,wu2020a}.  
The work of Wu, {\it et al.} examines the effect of a number of probability 
distributions for randomly assigned response thresholds.  
They find that, from among the distributions tested, 
a uniform distribution provides the best goal achievement for a collective 
control 2-D tracking problem \cite{wu2021} in which the swarm controls a tracker 
object with the goal of following a target object in real-time \cite{wu2020a}.  

Because the agents in a swarm work collectively
and because threshold assignments both across\footnote{Distribution
of threshold values for a single task across all agents.}
and within\footnote{Distribution of threshold values across tasks
within a single agent.}
agents affect collective behavior,
determining the optimal distribution and assignment of thresholds in a swarm
is a non-trivial task.
Even more challenging, task demands may change over time in many problems. 
For example, in the 2-D tracking problem referenced above,
the target may make frequent, random changes in direction.  
One way for the swarm to adapt to these changes
is through dynamism in response thresholds \cite{castello2013,theraulaz1998}.  
In practice, however, systems that use 
dynamic thresholds can have difficultly adapting to new task demands \cite{kazakova2018,meyer2015,theraulaz1998}.
Thus, we explore {\it a priori} evolution of thresholds rather than real-time adaptation.
Genetic algorithms (GAs) are known for their ability to find effective solutions to computationally expensive, 
high-dimensional problems \cite{forrest1993}.  

In this paper, we demonstrate that response thresholds evolved by a genetic algorithm not only outperform random 
response thresholds but in many cases achieve nearly perfect performance for the domain problem.  Further, 
evolved thresholds outperform dynamic thresholds despite the apparent advantage of dynamism: the ability to 
adapt in real-time to changing task demands.  Secondly, we show that thresholds evolved for some problem instances
generalize to other problem instances with very different task demands.  
Two of the problem instances that we examine generalize to all other problem instances tested, 
making them \textit{universal training instances}.  
Finally, we analyze generalization between problem
instances to understand not only the relevant features in universal training instances for this problem but also to
aid in identifying universal training instances for other domain problems.

%% file: sec-bg.tex
\section{Background}
\label{sec:bg}

Self-organizing swarms are both robust and adaptable.  Robustness stems from the absence of centralized control.
Adaptability is a result of the large behavior space created by individual agents' independent decisions.  This 
behavioral diversity is critical for effective division of labor \cite{ashby1958}.  In response
threshold models, individual decisions are determined by comparison of sensed stimuli with internal threshold values.
The thresholds may be global or individual.
Let $\tau$ be the stimulus and $\theta$ the threshold.  In the most direct form of response threshold model, an
agent activates for the task if $\tau \geq \theta$.  In more complex models, activation decisions may depend on
additional information such as past performance, the work of other individuals, and randomness.

Inspired by observations of insect societies, the probabilistic response threshold model uses the stimulus and threshold
values to define a probability of response \cite{bonabeau1996,bonabeau1998}.  The probability that agent $i$ 
activates for task $j$ is given by
\begin{equation}
P_{i,j} = \frac{\tau_j^2}{\tau_j^2 + \theta_{i,j}^2}
\end{equation}
This results in a response probability of $0.5$ when $\tau = \theta$.  The probability approaches 0 if $\tau \ll \theta$ and 1 if $\tau \gg \theta$.  Diversity of behavior is inherent in this model due to its stochastic nature.  Even two agents with
the same threshold may not respond in the same way.  This model has gained wide acceptance in the multi-agent 
systems community \cite{campos2000,castello2018,castello2013,cicirello2002,correll2008,delope2013,dossantos2009,goldingay2013,kittithreerapronchai2003,merkle2004,niccolini2010,nouyan2005,pang2017,price2004,yang2010}.
 
Eliminating stochasticity creates a determinisitic model in which agent activation is defined by
\begin{equation}
P_{i,j} = \begin{cases}
			1 \ \mbox{ if } \tau_j \geq \theta_{i,j} \\
			0 \ \mbox{ otherwise}
		\end{cases}
\end{equation}
This determinism allows for an understanding of agent activations that is not possible in probabilisitic models.  With homogeneous response thresholds in a deterministic response model with global stimuli, agents activate in lockstep.  Let  $\alpha_{i,j}$ be the activation count for agent $i$ for task $j$.  
Then $\alpha_{i,j} = \alpha_{k,j} \ \forall \ i,k$.  In deterministic response models, diversity in agent activations is achieved through heterogeneity in response thresholds \cite{campbell2011,dossantos2009,kanakia2016,krieger2000a,krieger2000b,riggs2012,wu2018}.  
Let $\theta_{i,j} < \theta_{k,j}$ for agents $i, k$.  Then $\alpha_{i,j} \geq \alpha_{k,j}$.  That is, an agent with a lower
threshold for a task will activate for that task at least as many times as an agent with a higher threshold for 
that task.  We note that if demand exceeds an agent's thresholds for more than one task, a secondary selection mechanism must be applied.

Task specialization is common in natural swarms \cite{jeanne1986} and desired in artificial swarms due to the resultant decrease in task switching.  The degree of specialization may be proportional to colony size  \cite{holbrook2011,holbrook2013}.
While specialization has obvious benefits, it may have costs as well.  For example, a specialist robot may remain idle for a long period of time as it searches for a task to undertake \cite{brutschy2012}.

As in natural swarms, artificial swarms can benefit from specialization due to costs of task switching such as energy consumption and time, and due to the positive effects of learning for some tasks.  In self-organized swarms, specialization must be designed into the system, developed through adaptation, or evolved by the system.  Evolution may take the form of a simple parameter adaptation to develop division of labor \cite{labella2006}.
In some cases, agent behaviors evolve specialization to the extreme case in which agents never switch tasks.  A more desirable scenario is one in which agents evolve to specialize but maintain a sufficient degree of behavioral plasticity to allow activation for different tasks \cite{tuci2014}.
CONE, combining neuro-evolution with co-evolution techniques, evolves controllers that improve collective performance through agent specialization for a pursuit-evasion problem and a collective construction task \cite{nitschke2012b,nitschke2012a}.
As the costs associated with task switching increase, groups may be more likely to evolve specialization \cite{goldsby2012,goldsby2010}.

Variability in response thresholds promotes specialization since agents with low threshold values for a task will activate more frequently, reducing task switching.  The probability distribution used to generate threshold values affects the degree of specialization \cite{wu2020a}.  In threshold-based systems, evolving threshold values provides another mechanism for self-organized division of labor \cite{duarte2012}.

There is evidence that for some insect species, experience on a task causes individuals to specialize for that task.  For example, some species are known to increase activation probability for a task with experience performing that task \cite{gautrais2002,langridge2004,ravary2007,theraulaz1998}.  For example, {\it Leptothorax albipennis} and {\it Cerapachys birori} ants are more likely to activate for tasks on which they have had past success.  This suggests that individuals adapt response thresholds with experience.  This form of adaptability has been implemented in numerous multi-agent systems \cite{campos2000,castello2018,castello2013,cicirello2003,delope2013,delope2015,gautrais2002,goldingay2013,kazakova2018,kazakova2020,price2004,theraulaz1998}.   

By adapting agent behavior to changing task demands, dynamic response thresholds appear to promise improved specialization and swarm performance.  In practice, however, agent thresholds often migrate to sink states due to implementations that act as positive feedback loops.  This achieves increased specialization but fails to deliver on adaptability \cite{kazakova2018,kazakova2020,theraulaz1998}.  Sink states may be determined by the first tasks for which agents activate \cite{meyer2015}.

Evolutionary computation has been used to evolve several aspects of artificial swarms.  One of the most common applications is evolving agent and swarm behaviors.   These include the use of grammatical evolution to evolve collective behaviors \cite{neupane2019,neupane2018}, evolution of behavioral rules for a swarm \cite{samarasinghe2018}, evolving diversity using a decentralized MAP-Elites algorithm \cite{hart2018}, evolving collaboration using cooperative coevolution \cite{panait2006}, evolving cooperative behaviors \cite{wang2019}, and evolving selfish behaviors \cite{yamada2013}.  Swarm size has also been examined with respect to evolving swarm behaviors \cite{fischer2018}.  

Evolution of swarm organizations and coalitions is also the subject of significant research \cite{aubert-kato2017,guo2020,yu2011}.  Population management, in response to changing threat levels, has been evolved as well \cite{beckman2008}.

In the domain of swarm robotics, evolution of control mechanisms is an active research area.  
Work in this area typically involves one of several forms of neuro-evolution.
Examples include evolving controllers for: self-organization for a swarm of {\it s-bots} \cite{dorigo2004}, aggregation behaviors \cite{baldassarre2003,gomes2013b,gomes2013a,soysal2007,trianni2003}, coordinated motion \cite{baldassarre2007,sperati2008},  collective behaviors of autonomous vehicles \cite{huang2017}, specialization for a robotic team undertaking a construction task \cite{nitschke2012a}, communication network formation \cite{hauert2009}, exploration and navigation \cite{sperati2011}, rough terrain navigation \cite{trianni2006}, transport problems \cite{gross2008}, agent communication \cite{ampatzis2008}, learning behaviors \cite{pini2008}, primitive behaviors triggered by a pre-programmed arbitrator \cite{duarte2014}, several behaviors for aquatic surface robots \cite{duarte2016a}, and intruder detection \cite{duarte2016b}.  In addition, researchers have explored the relationship between evolution and the environment \cite{steyven2017}.

Evolutionary computation is used to evolve agent-level parameters to elicit desirable swarm behavior \cite{narzisi2006} and agent-level ranking schemes for task selection \cite{moritz2015}. 

As noted above, division of labor is an important factor in swarm success.  In this too, evolution may play a role. 
In evolutionary robotics, collective neuro-evolution is used to evolve agent specialization for gathering and collective constuction problems \cite{nitschke2009, nitschke2012b, nitschke2012a}.  Division of labor has been evolved in groups of clonal organisms \cite{goldsby2010}.   Using grammatical evolution, specialization is evolved via task partitioning \cite{ferrante2015}.

Ana Duarte {\it et al.} evolve response thresholds for a probabilistic response-based swarm with two tasks \cite{duarte2012}.  Two different single-objective outcomes are examined: work distribution and specialization.  
Task stimuli increase by a constant in every timestep and decrease by the work performed.  The balance between task demands is fixed at either 1:1 or 3:1 throughout a run.  They find that threshold values evolve to 0 when work distribution is the evolutionary objective, leaving assignment of workers to tasks entirely dependent on stimulus parameters.  Specialization develops when costs associated with task switching are high. 
Our work differs from this in several significant ways.  First, we use a deterministic threshold-based system.  More importantly, our testbed allows exploration of task demands that are dynamic and highly variable requiring a much greater degree of adaptability from the swarm (see Section~\ref{sec:model} for details).  Finally, we explore the degree to which thresholds evolved for one set of task demands generalize to problem instances with very different task demands.

%% file: sec-model.tex
\section{Task Allocation in a Threshold-based System}
\label{sec:model}

The testbed we use in this work is a collective control problem in two 
dimensions.  
\begin{figure}
\centering
\includegraphics[width=0.65\textwidth]{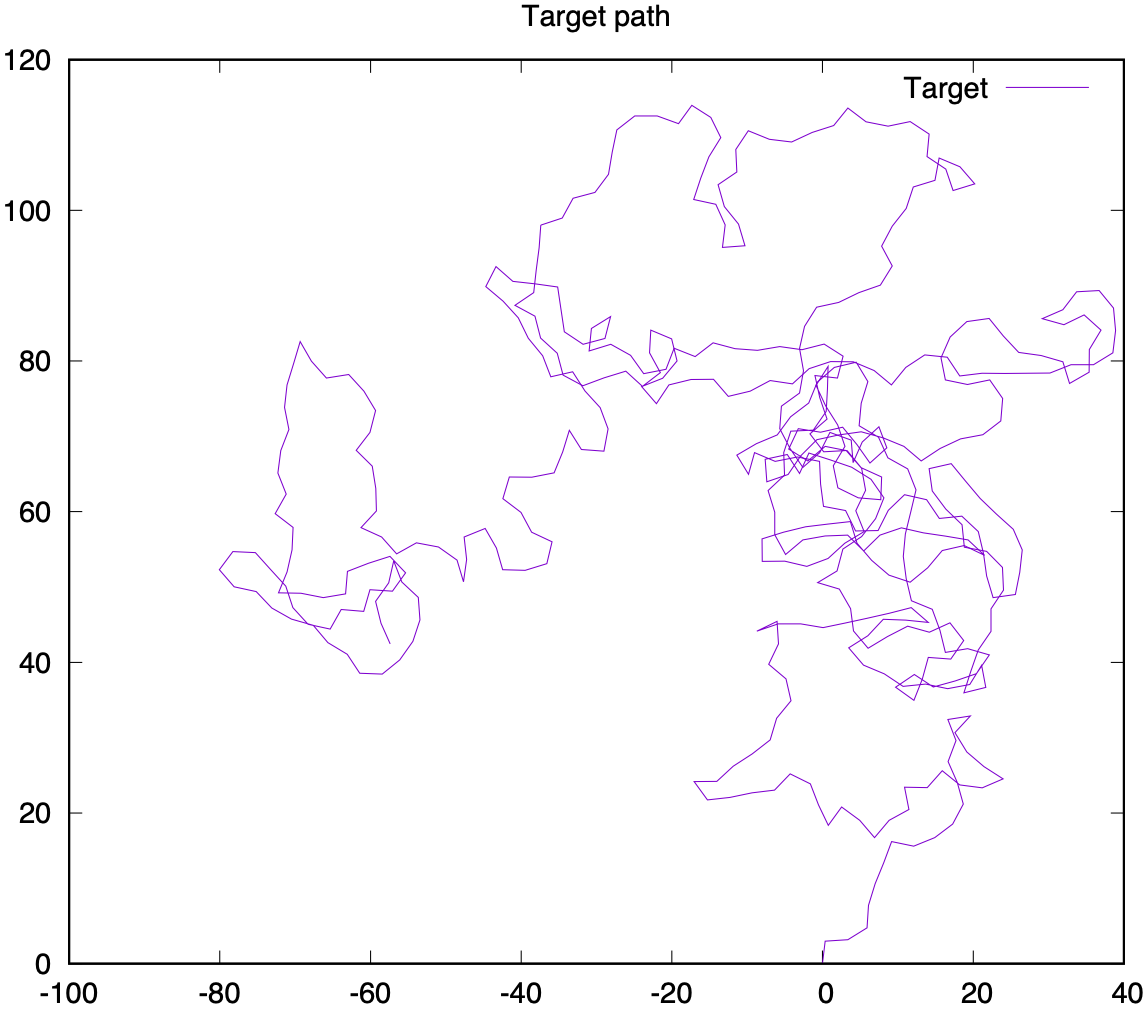}
\caption{An example \simpath{random} path with length 1500 representing 500 timesteps.}
\label{fig:path-random}
\end{figure}
Weidenm\"{u}ller describes the problem of honeybee nest thermoregulation, a one-dimensional collective control problem from nature \cite{weidenmuller2004}.  We increase the number of dimensions and cast the problem as a 2-D tracking problem \cite{wu2021} to allow more natural visualization of results.  We note, however, that the 
focus of this work is on achieving effective decentralized task allocation via inter-agent 
variation.  The experiments in this work would be largely unchanged if we cast the problem into a different domain, including those without a physical manifestation.  Thus, we abstract physical considerations out of the problem. 

In this problem, the swarm collectively pushes a tracker object, attempting to follow the target as closely as possible.  A simulation consists of a pre-determined number of timesteps.  In each of these timesteps, the target moves a fixed distance defined by parameter \texttt{target\_step\_len}.  Figure~\ref{fig:path-random} shows an example path in which the direction of movement in each timestep is determined at random.

Four tasks are defined: \texttt{push\_NORTH}, \texttt{push\_EAST}, \texttt{push\_SOUTH}, or \texttt{push\_WEST}.  
Each agent $i$ has a threshold 
$\theta_{i,\textsc{d}}$ 
for each $\textsc{D} \in \{\texttt{NORTH, EAST, SOUTH, WEST}\}$.
Thresholds are in $[0, 1]$ and are mapped to the domain space via the {\tt range} parameter.
Task demands are determined by the relative positions of the target and tracker.  Let $\Delta x = \mbox{target.}x - \mbox{tracker.}x$ and $\Delta y = \mbox{target.}y - \mbox{tracker.}y$.  Task stimuli are defined as: $\tau_N = - \Delta y$, $\tau_E = - \Delta x$, $\tau_S = \Delta y$, and $\tau_W = \Delta x$.  

The defining characteristic of our testbed is that by changing the target path we can dramatically alter the demand placed on the swarm with respect to task demands.  Demand can change gradually or abruptly, at a constant rate or a changing rate.  The target can move and turn in all directions or in only some.  A single problem instance can include many of these characteristics at once.  For example, in \simpath{zigzag} demand for \texttt{push\_EAST} is constant 
while, for \texttt{push\_NORTH} and \texttt{push\_SOUTH},
it is punctuated by abrupt changes.  There is no demand for \texttt{push\_WEST}.  The balance of demands among the tasks changes little.  In contrast, \simpath{circle} provides equal demand for all tasks over time, with continuous changes in demand and balance per timestep.  The variety of demands that can be placed on the swarm make this an ideal testbed for task allocation.

Our main results in this work are for evolved, static response thresholds.  We compare the results for evolved thresholds with earlier work, including the use of dynamic response thresholds.  In the model of dynamism used for comparison in this work, 
threshold values can vary in $[0, 1]$.  
$\theta_{i,\textsc{d}}$ 
decreases in each timestep during which agent $i$ activates for task $\textsc{D}$ and increases in each timestep during which the agent activates for another task \cite{wu2020ants}.  Threshold values are initialized uniformly at random in $[0, 1]$.

We evolve and test response thresholds for six target paths: \simpath{circle}, \simpath{diamond}, \simpath{random}, \simpath{scurve}, \simpath{square}, and \simpath{zigzag}.  Detailed descriptions appear below.
\begin{itemize}
\item \simpath{circle}: Target continuously revolves about a central point at a fixed distance, resulting in a circular path with radius $r$.  This creates continuously changing task demands and requires the swarm to perform all tasks equally over the course of a run.

\item \simpath{diamond}: Target moves continuously along the perimeter of a square rotated 45 degrees from the axes.  Thus, all motion creates simultaneous demand for two tasks.  Size is determined by parameter \texttt{Edge\_length}.

\item \simpath{random}: In each time step, target direction is calculated from the current heading by adding an angle, in radians, drawn from the Gaussian distribution $\mathcal{N}(0.0, 1.0)$.  With high probability, random requires all tasks.  See Figure ~\ref{fig:path-random}.

\item \simpath{scurve}: A periodic, rounded path that oscillates up and down, moving from west to east.  The motion is defined by parameters {\tt Path\_amplitude} and {\tt Path\_period}.  Requires all tasks, however, there is very little demand for \texttt{push\_WEST}.  See Figure ~\ref{fig:scurve-zigzag}.
\begin{figure}[t]
\centering
\includegraphics[width=0.85\textwidth]{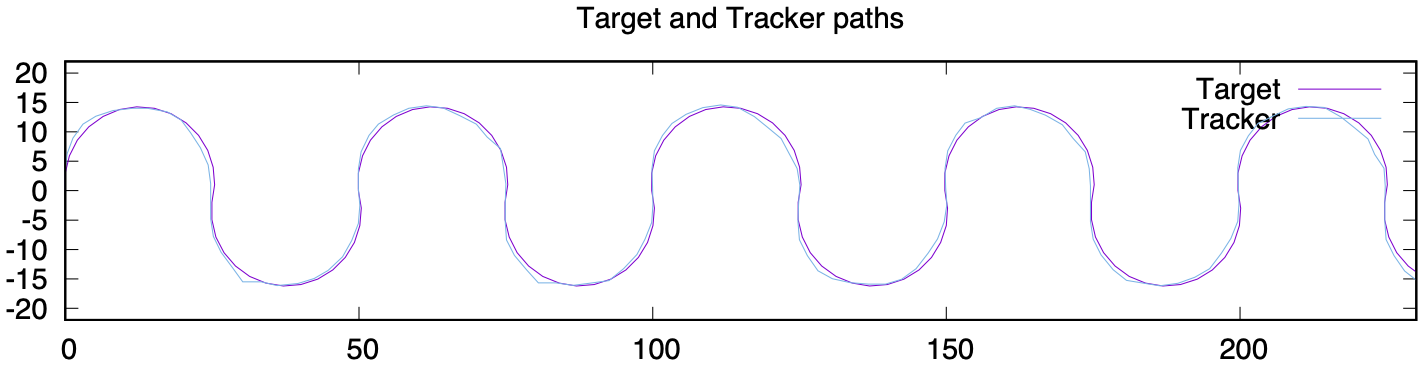} \\ [2mm]

\includegraphics[width=0.85\textwidth]{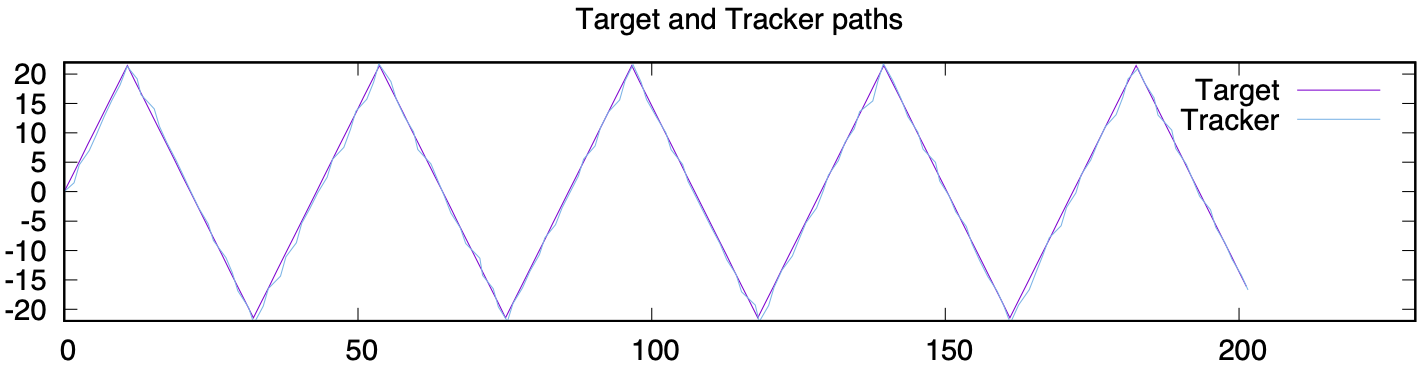}
\caption{Partial examples of scurve (top) and zigzag (bottom).}
\label{fig:scurve-zigzag}
\end{figure}

\item \simpath{square}: Target continuously moves along the perimeter of an axis-aligned square with edge length defined by parameter {\tt Edge\_length}.  Requires all tasks during a run but motion at any given time 
is in only one direction.

\item \simpath{zigzag}: A periodic, oscillating path moving from west to east with identical, angular straight edges moving alternately northeast and southeast.  As for \simpath{scurve}, amplitude and period are parameterized.  No demand for \texttt{push\_WEST} but all motion creates simultaneous demand in two dimensions.   See Figure ~\ref{fig:scurve-zigzag}.
\end{itemize}

In each timestep, each agent activates for one task or remains idle.  Candidate tasks for agent $i$ are those for which global stimulus $\tau_{\textsc{d}} \geq \theta_{i,\textsc{d}}$.  If there is more than one candidate task in a timestep, selection is random.

Performance is evaluated according to two 
domain goals:
\begin{goal} \label{avg_diff}
Minimize the average positional difference, per time step, between the target location and the tracker location. 
\end{goal}
\begin{goal} \label{total_dist}
Minimize the difference between total distance traveled by target and the total distance traveled by the tracker.
\end{goal}
It is important to note that both domain criteria are necessary to gauge the swarm's success.  If using only Goal \ref{avg_diff},  the tracker could remain close to the target while zigzagging repeatedly across the target path.  This would yield strong performance with respect average positional difference but a path length that is significantly greater than that of the target.  Alternately, using only Goal \ref{total_dist}, the tracker might travel a path that is the same length as that of the target but unrelated with respect to shape and direction.

We recognize that there are more effective solutions to the tracking problem.
It is not our goal to find the most efficient method to solve this problem.
Instead, the collective tracking problem 
serves as our testbed because it is a decentralized task allocation problem in 
which task demands are clearly defined and measured, 
dynamic variation in task demand over time can be systematically described,
and overall performance can be accurately measured as well as visually assessed.  
We reiterate that our focus is on decentralized, dynamic task allocation
via inter-agent variation in general and variation in response thresholds in particular.
Modeling physics, or physical robots, is not the purpose of this work.

%% file: sec-ga.tex
\section{The Genetic Algorithm}
\label{sec:ga}

Our genetic algorithm is based on NSGA-II \cite{nsga-ii} and implemented in Python using the Distributed Evolutionary Algorithms in Python (DEAP) library \cite{fortin2012}.  Though there are newer multi-objective genetic algorithms, NSGA-II is well understood, widely used, and very effective for small numbers of objectives.  
We use three optimization objectives: 
\begin{itemize}
\item minimize average positional difference: the average, over all timesteps in a simulation, of the distance between the target and tracker 
\item minimize path length difference: the difference in total distance traveled by the target and tracker 
\item minimize average number of task switches: the average, over all agents, of the number of switches from one task to another
\end{itemize}  
The first two objectives are aligned with the domain goals above.  The third is a measure of agent specialization.  Each individual in the GA population represents a complete swarm consisting of 50 agents.

\begin{figure}
\centering
\includegraphics[width=0.5\textwidth]{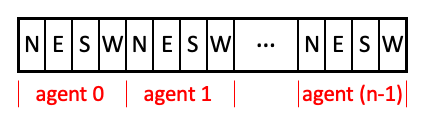}
\caption{Structure of our genome.  It consists of a threshold for each of the tasks for each agent.  The thresholds are real-valued.}
\label{fig:genome}
\end{figure}

Illustrated in Figure~\ref{fig:genome}, the genome consists of 200 real-valued numbers, four for each of the 50 agents represented by an individual.
The values for an agent represent the four task thresholds for that agent, each a real value in $[0, 1]$.
Thus, in aggregate the genome is:
$[\theta_{i,\textsc{d}}], \ \forall \ \textsc{D} \in \{{\texttt{NORTH, EAST, SOUTH, WEST}}\} \ \mbox{and} \ \forall \ i \in [0,49]$.  Individuals are initialized with threshold values generated uniformly at random in $[0, 1]$.  

In each generation, we create an offspring population that is initially a copy of the parent population.
Offspring are randomly paired for crossover, which occurs with probability $0.9$.  The crossover operator is uniform, with individual threshold values exchanged between individuals with probability $0.7$.  
Individuals that undergo crossover are mutated with probability $0.2$ while those that do not are mutated with probability $1.0$.  In the mutation operator, each agent within a GA individual is mutated with probability $2/m$, where $m$ is the number of agents.  An agent mutation consists of choosing one threshold value uniformly at random and 
adding a small $\Delta$ generated uniformly at random in a parameterized range centered at 0.  After crossover and mutation are complete, the offspring are combined with
the parent population.  Replacement is via the standard NSGA-II operator.

Fitness is determined by running the simulator using the threshold values for an individual and the objectives defined above.  For deterministic paths,
evaluation consists of a single simulation while for random paths we run three simulations and average the results.


%% file: sec-results.tex
\section{Experiments and Results}
\label{sec:results}

For each target path, we perform 32 runs of the genetic algorithm and choose one individual from front 0 of each run.  We use these individuals for testing on the path used for evolution as well as all other paths.  For each test, we perform 30 runs of the simulation, averaging positional difference, path length difference and number of task switches across the runs.  Table ~\ref{tab:params} lists the GA and simulator parameters used in our
experiments.

\begin{table}
\centering
\begin{tabular}{l r}
\hline
Simulation Parameter  \hspace{0.75in} & Value \\
\hline \hline
Population size & 50 \\
Simulation timesteps, training & 200 \\
Simulation timesteps, testing & 500 \\
Task selection & random \\
Target\_step\_len & 3 \\
Step\_ratio & 2.0 \\
\hline
 \\
GA Parameter \hspace{0.75in} & Value \\
\hline \hline
Population size & 100 \\
Initialization & $\mathcal{U}(0.0, 1.0)$ \\
Mutation $\Delta$ & $\mathcal{U}(-0.1, 0.1)$ \\
Mutation expected changes & 2 \\
Selection &  random \\
Crossover &  uniform \\
Crossover probability & 0.9 \\
Uniform probability & 0.7 \\
Replacement &  NSGA-II \\
Generations & 2500 \\
\hline
\end{tabular}
\caption{Simulation parameters (top) and GA parameters (bottom).  Values for Target\_step\_len and 
Step\_ratio are used to allow comparison with previous results.}
\label{tab:params}
\end{table}

Individuals in the GA population are evaluated by running swarm simulations.  This is performed for each unevaluated
offspring in each generation, up to 100 individuals per generation.  For a deterministic path, this results in up to 250000
simulations for a GA run.  To reduce the time required for these runs, we use only 200 timesteps per simulation during GA runs.  During testing of the evolved thresholds, we use 500 timesteps per simulation to allow comparison with 
previous results in which we use 500 timesteps.

In deterministic threshold-based swarms, uniformly distributed thresholds provide better swarm performance
than do thresholds generated by Gaussian, Poisson, or logarithmic distributions for a tracking problem like that used here \cite{wu2020a}.  In this section, we provide experimental results demonstrating that evolved response thresholds outperform uniform thresholds and dynamic thresholds with respect to tracking metrics.  In addition, 
thresholds evolved for one path generalize, to varying degrees, to other paths.  We 
identify a class of paths for which our algorithm evolves universal response thresholds that provide nearly optimal performance for all other tested paths.  Finally, we analyze these universal training instances to identify
features that make universality possible.

\subsection{Effect of Evolved Thresholds}
\label{sec:evolved_effect}

\begin{figure}
\centering
\includegraphics[width=0.7\textwidth]{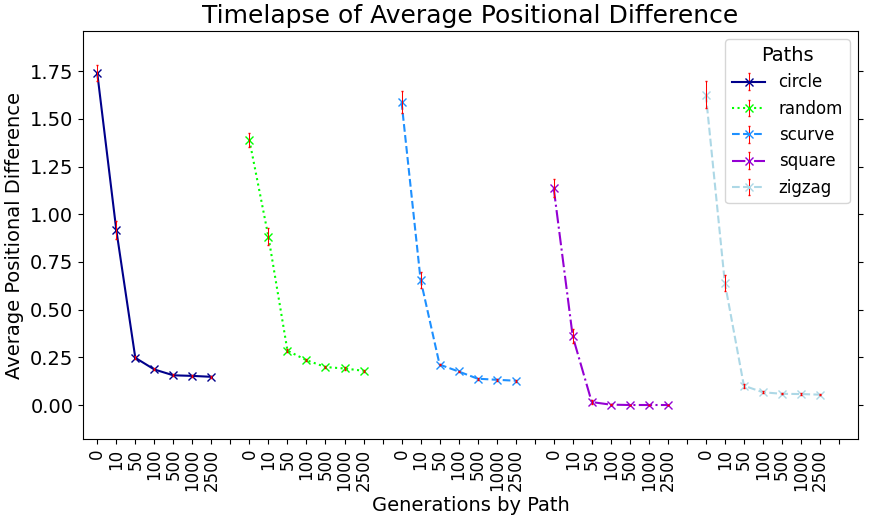} \\
\includegraphics[width=0.7\textwidth]{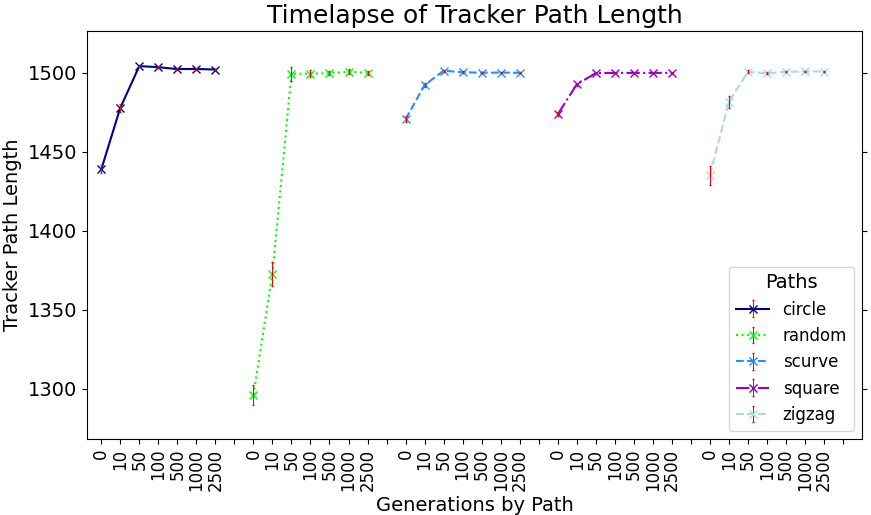}
\caption{Convergence of threshold evolution by target path.  Each data point represents one simulation run
for each of the 32 GA runs.  95\% confidence intervals are shown in red.  For each series, generation 0 represents thresholds generated uniformly at random.}
\label{fig:convergence}
\end{figure}

Figure ~\ref{fig:convergence} illustrates the effect of evolved thresholds on swarm performance for the 2-D tracking problem.  
Each plot shows data for each of the six paths tested.  The x-axes show generations of evolution for each of the paths.  In the top plot, the y-axis represents the average positional difference between target and tracker.  The y-axis in the bottom plot represents the tracker path length.  Target path length for the runs shown is 1500.  Each data point represents 32 simulations, one for each run of the genetic algorithm for that path.  Each run was performed on an individual chosen from front 0 after performing a non-dominated sort of the population.  95\% confidence intervals are shown in red.

Recall that the thresholds of the initial population in the genetic algorithm are uniformly distributed at random.  Thus, even at generation 0, the swarm performs well.  Evolved thresholds significantly improve performance with respect to both domain goals for all target paths.  For tracker path length, performance is nearly optimal for all paths.

\begin{figure}
\centering
\includegraphics[width=0.98\textwidth]{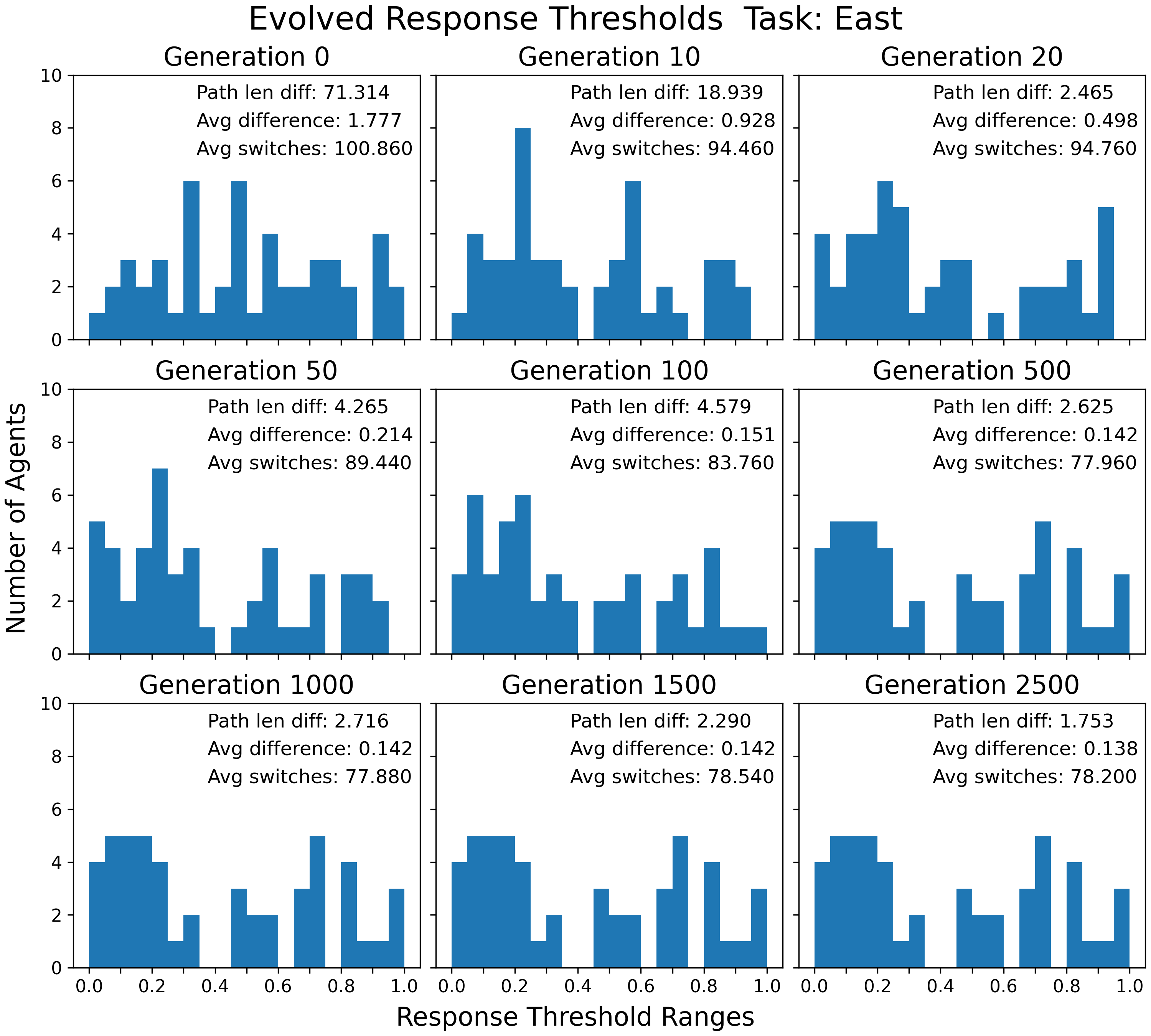}
\caption{Evolution of thresholds.  Initial thresholds are uniformly distributed at random.  The histogram shows 20 buckets of uniform width.  The GA is run for 2500 generations, though for the run shown the histogram does not change further after generation 500.  Threshold values may continue to evolve though the count in each bucket does not change.}
\label{fig:timelapse}
\end{figure}

Figure ~\ref{fig:timelapse} shows a series of histograms depicting the threshold values for task {\tt push\_EAST} at 
nine times during a run of the genetic algorithm for target path \simpath{circle} with radius 10.  
On the x-axis, the threshold range of $[0.0, 1.0]$ is divided into 20 buckets of width $0.05$.  The y-axis is the count of agents in each bucket.  The swarm consists of 50 agents.  In generation 0, the thresholds are the initial values, generated uniformly at random.  In each subplot, the swarm's performance using the current threshold values is shown at the upper right.  The individuals depicted are selected from front 0 after performing a non-dominated sort.

This sequence shows that the number of low thresholds very quickly increases, resulting in more agents activating for this task.  For the GA run shown, the histogram for {\tt push\_EAST} is fixed by generation 500, though values within the buckets may continue to change.

Examining activation counts for agents with the final thresholds reveals that only 25 of the agents activate for {\tt push\_EAST}.  In the plot for generation 500 in Figure ~\ref{fig:timelapse}, twenty-four agents are represented in the block with thresholds between 0.0 and 0.3.  All of these agents, plus one of the agents in range 0.3 to 0.35 (with only 10 activations), activate at least once.  None of the remaining agents activate for this task, {\tt push\_EAST}.  
This means that the thresholds for {\tt push\_EAST} for these agents constitute, in effect, a non-coding region of the genome as they do not affect fitness, unless the thresholds mutate sufficiently for that agent to activate.

\begin{figure}
\centering
\includegraphics[width=0.9\textwidth]{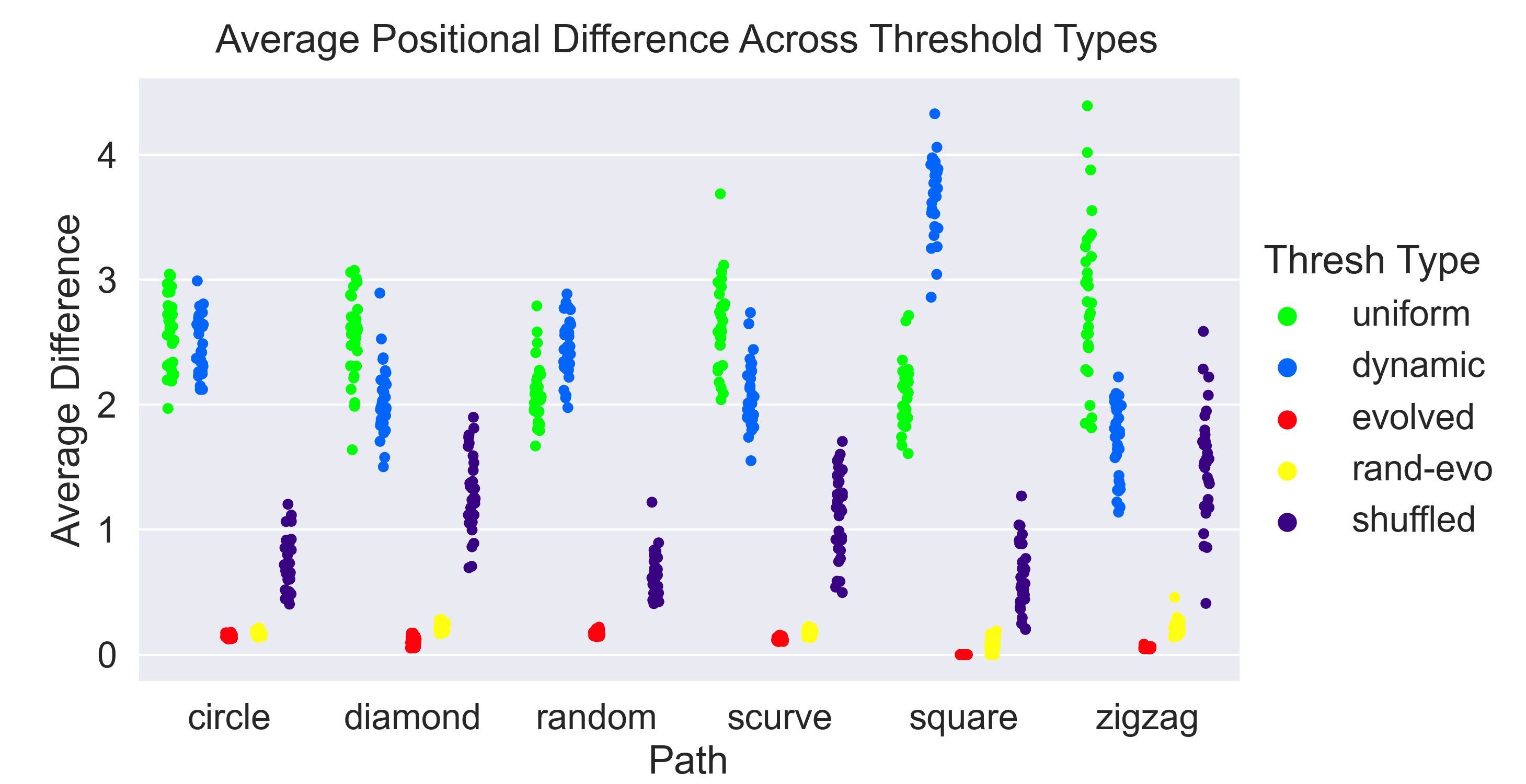}
\caption{Comparison of swarm performance for six problem instances (paths) using response thresholds generated in five ways: distributed uniformly at random, dynamically adjusted during a run, evolved by a GA for the test path, evolved by a GA for the \simpath{random} path, and shuffled.  Shuffled thresholds are evolved for the test path but are then randomly permuted across agents for each task.  This breaks the evolved balance of thresholds for each agent.  Shuffled thresholds are discussed further in Section ~\ref{sec:evolved_effect}.}
\label{fig:thresh-types}
\end{figure}

Introducing dynamism in response thresholds may allow a swarm to adapt to changing task demands and different
problem instances.  Because we evolve response thresholds for particular problem instances, it is natural to 
compare evolved thresholds to dynamic thresholds to determine if the expected benefit of dynamism is observed.  Figure ~\ref{fig:thresh-types} illustrates this comparison.  
The x-axis consists of six groups, one for each target path.  Within each group, we show data points for uniform thresholds, 
dynamic thresholds, thresholds evolved for that target path, and thresholds evolved for \simpath{random} paths but 
tested on the target path.  In the group for \simpath{random}, we omit the fourth data since they duplicate the third.  
The y-axis  represents average positional difference.  Each column represents 30 runs of the simulation for each 
of the 32 runs of the genetic algorithm.  Results for tracker path length are similar.

These results show that evolved thresholds significantly outperform both uniform thresholds and dynamic thresholds.
Dynamism is beneficial for some paths but not for others.  This may be due to the effects of the positive 
feedback loop on dynamic thresholds, causing them to migrate to sink states.  One possible advantage of dynamic
over evolved thresholds is that they adapt in real-time to any path, perhaps making them more general.  The rand-evo data demonstrate that this is not the case.  Using thresholds evolved for \simpath{random}
we find excellent performance for all paths suggesting that evolved thresholds can generalize.

Evolved thresholds also have a beneficial impact on specialization for 
most paths.  
For each of the 32 runs of the GA, we perform 30 swarm simulations using the evolved response thresholds.  Figure 7 shows the average task switches for these simulations for all six problem instances. 
\begin{figure}
\centering
\includegraphics[width=0.9\textwidth]{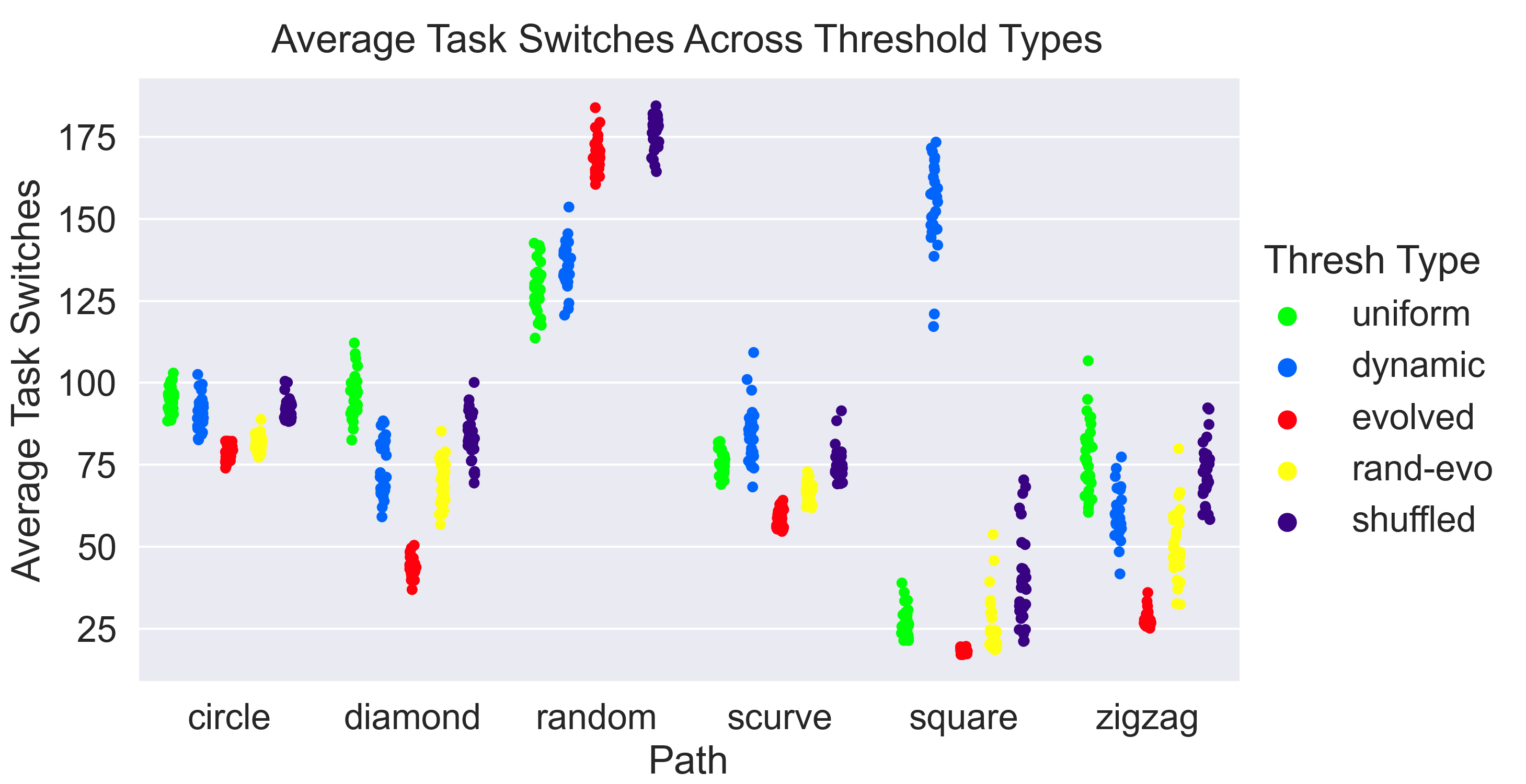}
\caption{Comparison of task switches for six problem instances (paths) using response thresholds generated in five ways.}
\label{fig:switches}
\end{figure}
Improvement ranges from 
modest for \simpath{square}, for which specialization is unnecessary due to demand for only one task in any timestep, to substantial for \simpath{diamond} and \simpath{zigzag}, which share the properties of long periods of unchanging task demand and demand for multiple tasks in all timesteps.  \simpath{random} is an outlier with a high number of task switches for evolved thresholds.  We hypothesize that this is due to each evaluation of individuals during evolution using different \simpath{random} paths.  Thus, while the balance of thresholds for each task is well-developed, as evidenced in Figure~\ref{fig:thresh-types}, the balance across tasks for each agent is not.

To gauge the effect of the relative values of thresholds for an agent, we randomly shuffle the evolved thresholds for all agents.  This maintains the distribution of threshold values for a task but disrupts the relative values across tasks for each agent.  We shuffle the evolved values for all 32 GA runs for each of the six paths.  We then perform 30 simulations with each set of shuffled thresholds.  The results are represented by the ``shuffled" data series in Figure ~\ref{fig:switches}.  With shuffled thresholds, specialization is dramatically reduced compared to unshuffled evolved thresholds for all paths except \simpath{random}.  For this path, shuffled and unshuffled thresholds exhibit similar specialization.

Note that this is not the case for the domain goals of average positional difference (Figure~\ref{fig:thresh-types}) and tracker path length.  For the domain goals, evolved thresholds perform well for \simpath{random}.  
The figure shows that shuffled thresholds perform better than uniform thresholds due to the improved distribution of thresholds for a task.  They are not, however, as effective as the unshuffled evolved thresholds due to the lack of coordination of thresholds for each agent.


\subsection{Generalizing Across Paths}

A reasonable concern for {\it a priori} evolution of response thresholds is that the result will be specific to the problem instance used during evolution.  If this is the case, evolution of thresholds would be required for each problem instance to be solved, a time-consuming undertaking.  We investigate this by testing the generalization of evolved thresholds across problem instances with greatly varying task demands.  We show that response thresholds evolved for some problem instances generalize well to all other instances.  By performing additional experiments, we attempt to identify the instance properties that facilitate generalization.

We have demonstrated that evolved thresholds result in better swarm performance, for the path for which they were evolved, than uniformly distributed thresholds.  Uniformly distributed thresholds generalize well.  That is, they perform well for all paths.  This is not surprising since uniformly distributed thresholds are general by definition.  It is not obvious that thresholds evolved for the task demands of one path should perform well for another path with different demands.  In this section, we show that evolved thresholds can generalize and that some paths result in thresholds that provide excellent performance for all other paths.

\begin{figure}[t]
\centering
\includegraphics[width=0.98\textwidth]{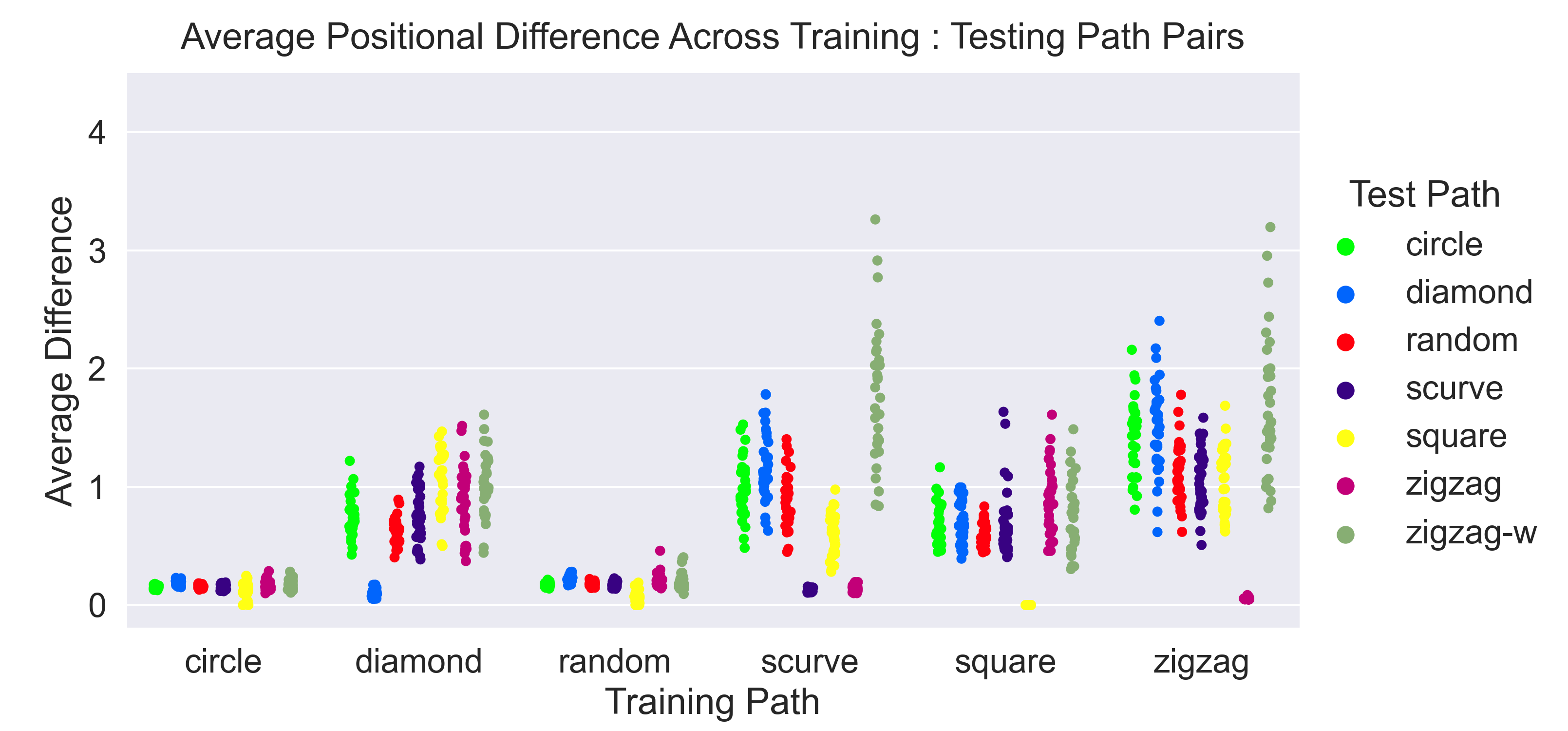}
\caption{Generalization of thresholds evolved for six paths.  \simpath{circle} and \simpath{random} provide the best generalization.}
\label{fig:generalize}
\end{figure}

Figure ~\ref{fig:generalize} depicts the degree of generalization for evolved thresholds.  
The x-axis labels represent the path used for evolution.  Each color represents a path used for testing.
Each column represents 32 data points, one for each run of the genetic algorithm.  Each data point is the average of 30 simulations using one set of evolved thresholds.  The y-axis shows average positional difference.

The figure shows that \simpath{circle} and \simpath{random} generalize to all other paths.  \simpath{circle} 
does slightly better than \simpath{random} for all paths except \simpath{square}.  Neither \simpath{square} nor 
\simpath{zigzag} generalize well, though we note that the performance is approximately the same as 
uniform thresholds for the testing path.  \simpath{scurve} generalizes well to \simpath{zigzag} but not to other paths.
The trend is the same for tracker path length though the plot is omitted for space considerations.

\begin{figure}[t]
\centering
\includegraphics[width=0.6\textwidth]{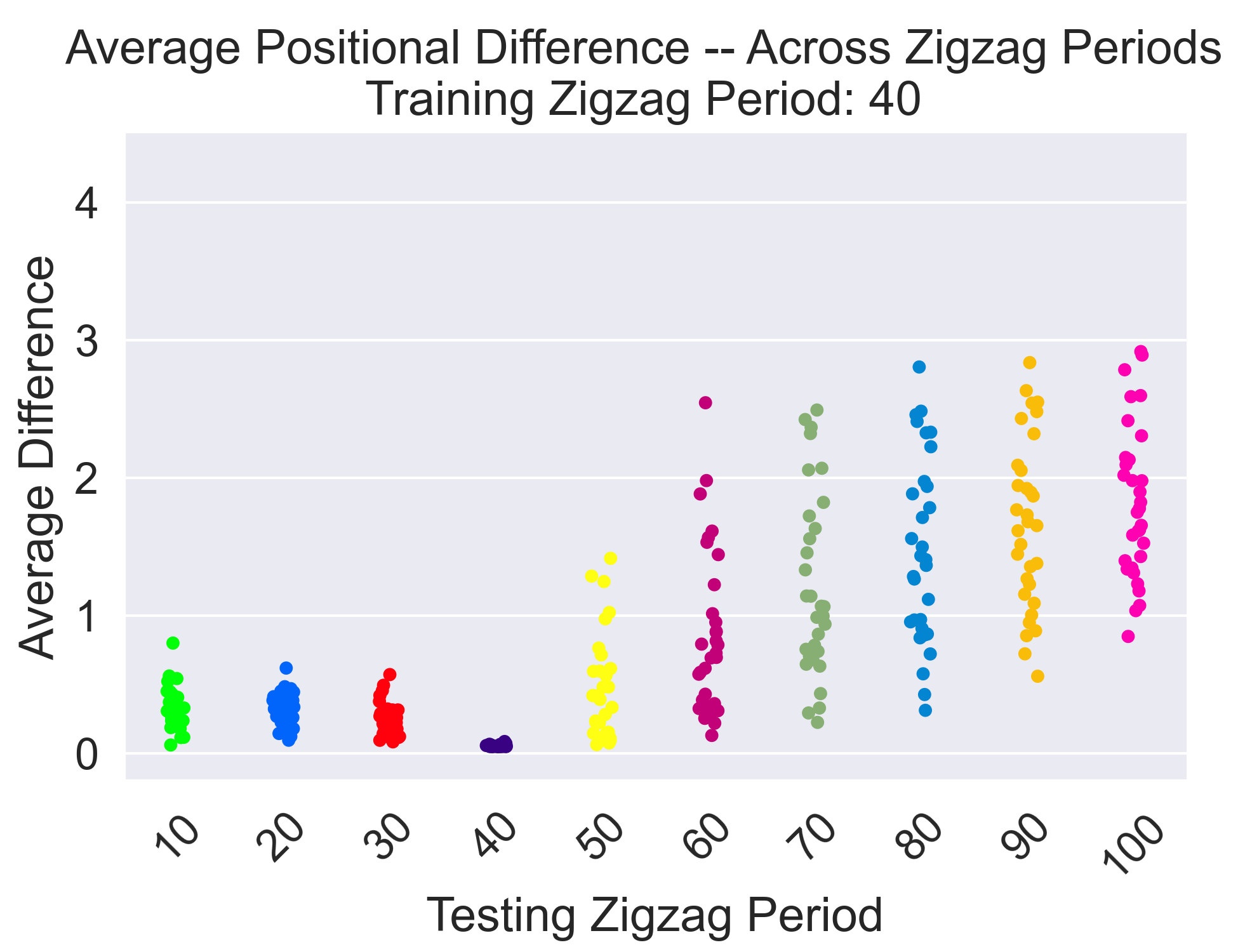}
\caption{Generalization results for thresholds evolved for zigzag with period 40, tested on
zigzag with periods of 10 to 100.}
\label{fig:zigzag-periods}
\end{figure}

\begin{figure}[t]
\centering
\includegraphics[width=0.6\textwidth]{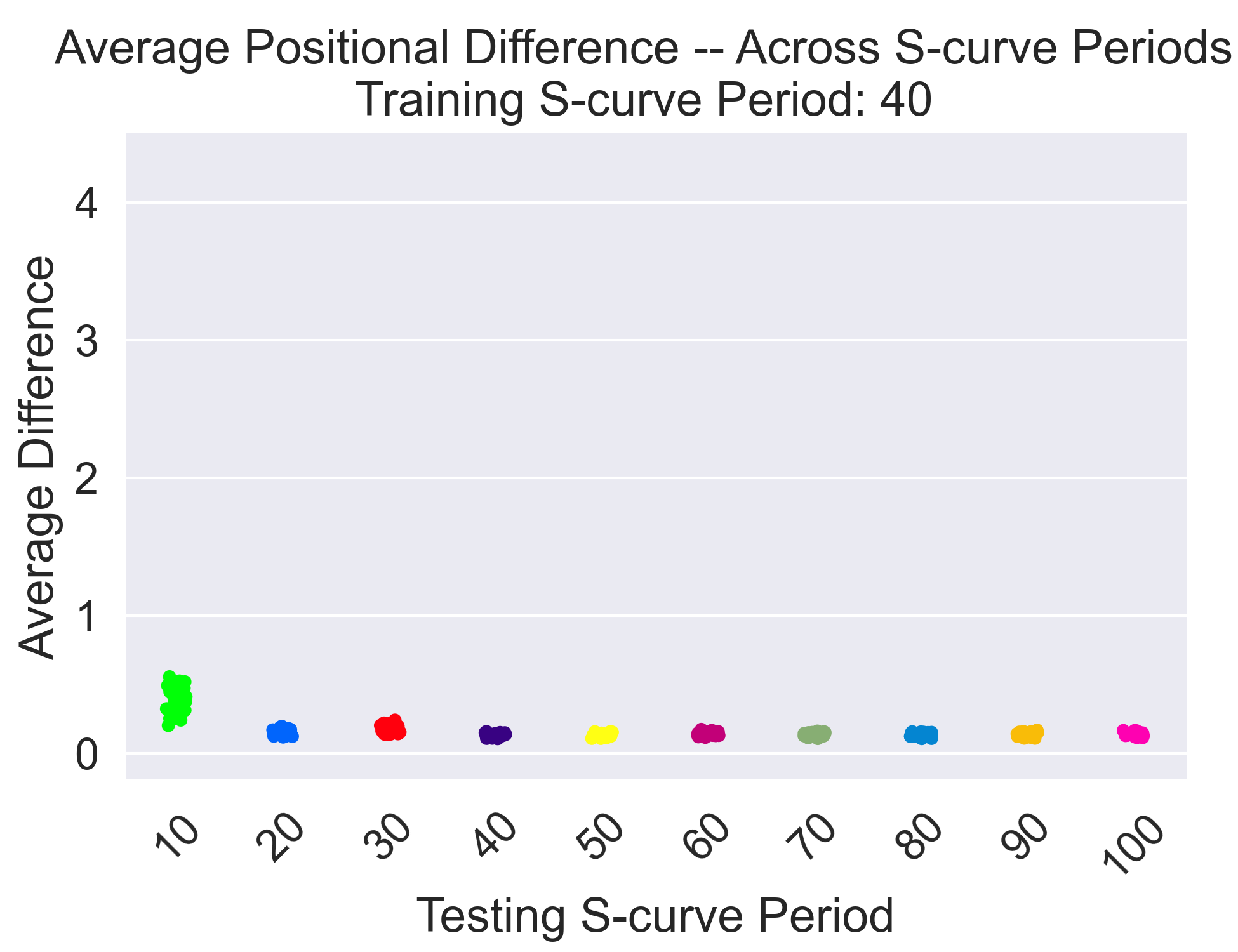}
\caption{Generalization results for thresholds evolved for scurve with period 40, tested on
scurve with periods of 10 to 100.}
\label{fig:scurve-periods}
\end{figure}

The first step in understanding these results is to revisit the path descriptions.  For \simpath{zigzag}, 
movement alternates periodically between straight lines to the northeast and southeast.  
Though inspired by a sine wave, \simpath{scurve} is somewhat more bulbous, creating continuously 
changing task demands.  This explains the asymmetry in generalization between these two paths.  
\simpath{zigzag} thresholds are not able to address the changes in task demands created by 
\simpath{scurve}.  We hypothesize that thresholds evolved for \simpath{zigzag} specialize
to the balance of task demands presented by a path that follows only two slopes.  To test this hypothesis,
we use thresholds evolved for \simpath{zigzag} with period 40 on runs for other \simpath{zigzag} instances 
with different periods.  Changing the period while keeping the amplitude fixed, changes the slopes of the
\simpath{zigzag} path segments.  As seen in Figure~\ref{fig:zigzag-periods}, this minor change significantly degrades performance.  
In contrast, \simpath{scurve} thresholds work well for \simpath{zigzag} as they are not specialized to 
a particular balance of task demands.  Figure~\ref{fig:scurve-periods} shows testing of \simpath{scurve} thresholds evolved for period 40
on \simpath{scurve} instances with periods from 10 to 100.  
There is very little change in performance across 
period values.

Neither \simpath{scurve} nor \simpath{zigzag} generalize well to \simpath{circle}, \simpath{random}, or 
\simpath{square}.  This is because \simpath{zigzag} creates no demand for task {\tt push\_WEST} and 
\simpath{scurve} creates very little.  Therefore, west thresholds do not significantly affect individual
fitness during evolution creating very little evolutionary pressure on these values.  As a result, 
\simpath{scurve} and \simpath{zigzag} thresholds' ability to satisfy {\tt push\_WEST} demand is no 
better than chance.  The last column in each group in Figure ~\ref{fig:generalize} supports this claim.  They represent 
tests performed on \simpath{zigzag-w}, a version of \simpath{zigzag} in which the major direction of travel has been 
reversed to west but is identical to \simpath{zigzag} in all other respects.  \simpath{circle} and \simpath{random} 
generalize to \simpath{zigzag-w} with almost identical results as those for \simpath{zigzag}.  Neither \simpath{scurve} 
nor \simpath{zigzag} generalize to \simpath{zigzag-w}, supporting our claim.

\simpath{square} requires further analysis.  {\tt push\_WEST} comprises one-quarter of the task demand for \simpath{square}, yet it does not generalize as well as \simpath{circle} and \simpath{random} which have similar west demand as a fraction of total task demand.  We hypothesize that this results from task demand for \simpath{square} being limited to one direction in any timestep.  Thus, in terms of swarm performance, there is no need for specialization as there is only ever one task to undertake at any time.  Thresholds evolved in this environment, therefore, do not perform well when demand for multiple tasks exists.

We present results for \simpath{diamond} to add additional support for our hypotheses regarding the experimental results.  Recall that \simpath{diamond} is simply \simpath{square} rotated 45 degrees so that the corners are aligned with the four cardinal directions.  Therefore, like \simpath{zigzag}, \simpath{diamond} consists of straight segments each of which creates demand for two tasks but like \simpath{square} it creates demand for all four tasks during the course of a simulation.  Results for \simpath{diamond} are similar to those for \simpath{square} but slightly worse for most target paths.  As with \simpath{zigzag}, \simpath{diamond} evolves thresholds with a fixed balance between tasks for which there are simultaneous demands.  \simpath{square} does not suffer from this effect.  This conclusion is reinforced by the similar performance for \simpath{diamond} and \simpath{zigzag} for test path \simpath{square} despite the apparent advantage for \simpath{diamond} due to demand for \texttt{push\_WEST}.  That advantage is apparent in the performance for test path \simpath{zigzag-w} (Figure ~\ref{fig:generalize}).

\begin{figure}[t]
\centering
\includegraphics[width=0.6\textwidth]{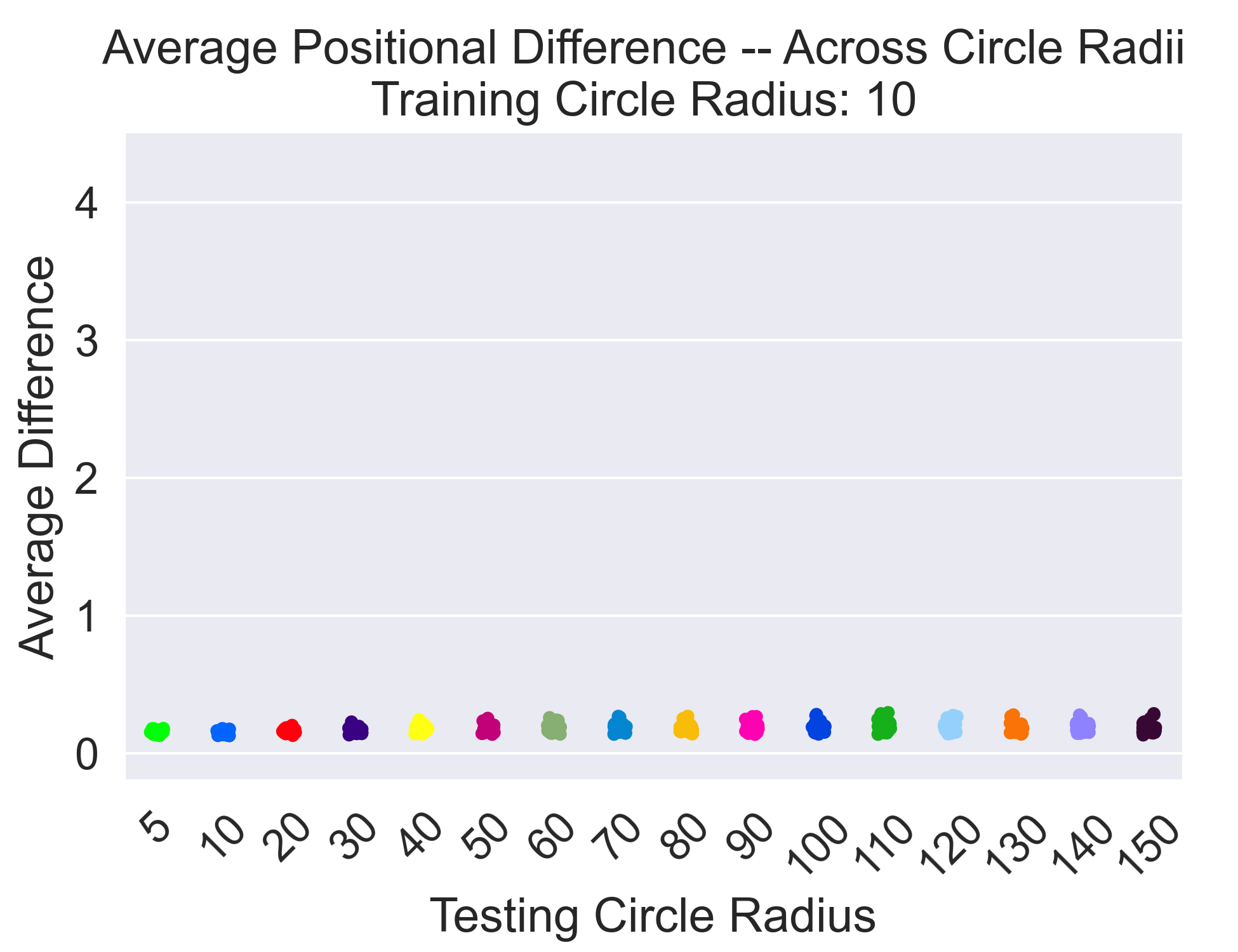}
\caption{Generalization results for thresholds evolved with a circle with radius 10, tested on
circles with radii of 5 to 150.}
\label{fig:circle-radii}
\end{figure}

\begin{figure}[t]
\centering
\includegraphics[width=0.6\textwidth]{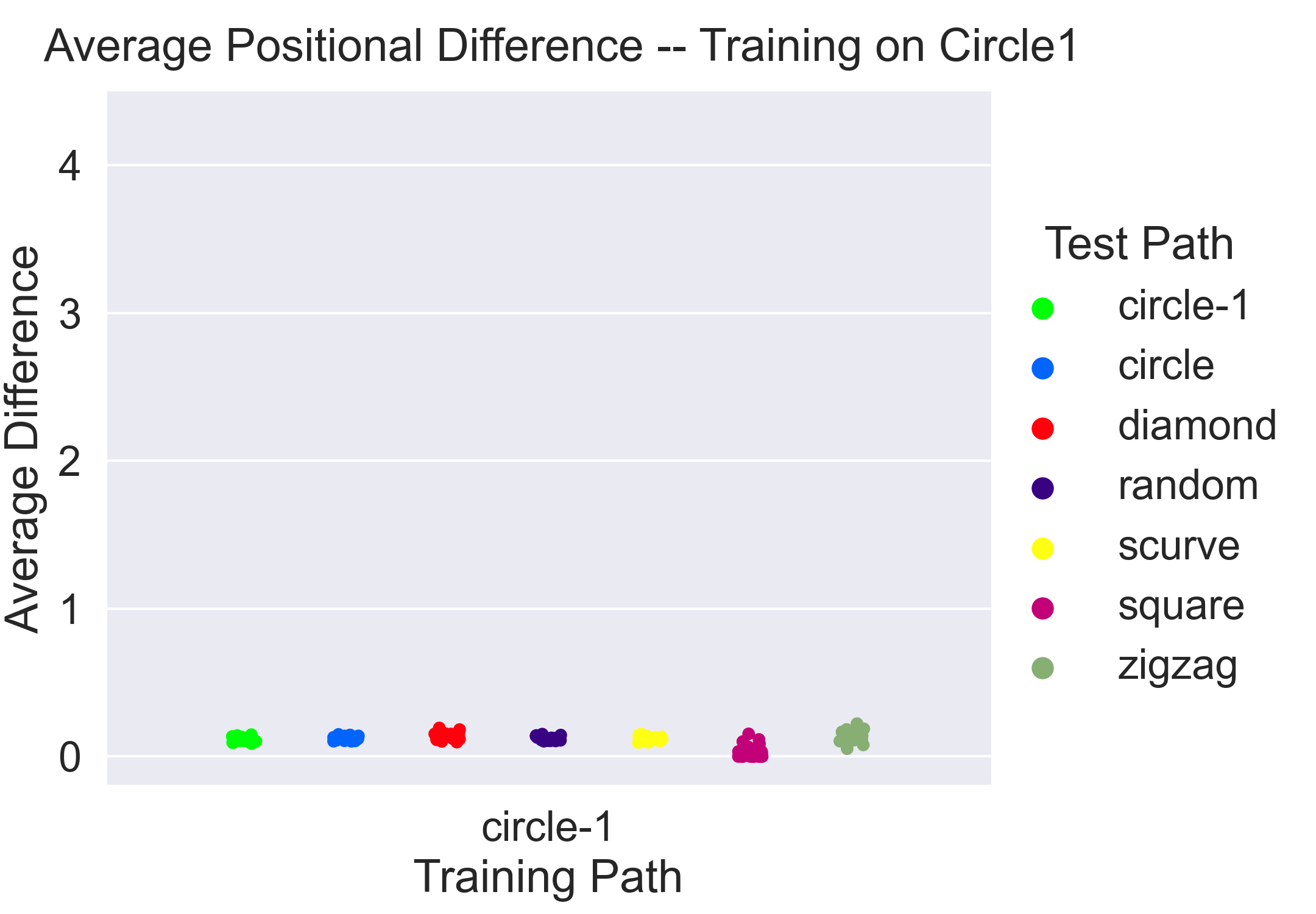}
\caption{Generalization results for thresholds evolved with fitness determined by only one revolution of a \simpath{circle}.  The y-axis matches the previous plots to facilitate comparison.}
\label{fig:circle-1}
\end{figure}

\simpath{circle} serves as a particularly robust training instance.  As shown in Figure ~\ref{fig:generalize}, \simpath{circle} thresholds generalize to all other paths.  In addition, they are effective for \simpath{circle} instances independent of radius.  A radius of 10
is used for evolving thresholds.  As Figure~\ref{fig:circle-radii} shows, tests on \simpath{circle} with radii from 5 to 150 show no significant change in performance.  
Further, thresholds evolved via GA runs in which fitness evaluation uses simulations making as little as one revolution of the \simpath{circle} results in values with the same performance as simulations making multiple revolutions.  Figure ~\ref{fig:circle-1} provides these results.  Using only 0.75 revolutions results in only slight degradation.

These results suggest two necessary and sufficient features for a universal training instance: sufficient demand for all tasks, and a wide range of simultaneous demand levels to allow thresholds to evolve to reasonably address any balance of demand between multiple tasks.  It is important to note that this does not require that the training instance includes every possible turn or curve that might be encountered in testing.  A simple circle in which demand changes in the same way, from one timestep to the next throughout a run, is universal.  There are no left turns, no sharp turns and no straight lines.  The number of timesteps during which there is demand for only one task is extremely small as these occur only when the target and tracker are at the same $x$ or $y$ location but are not co-located.  
Though a very different path, \simpath{random} provides similar task demands with respect to variety in the balance of simultaneous demand for multiple tasks and a high probability of all tasks being represented.
These results provide hope that simple universal training instances may exist for other problem domains.

\subsection{Threshold Distribution}
\label{sec:success}
Two factors explain the success of evolved thresholds: the distribution of threshold values for a task and the relative values of the thresholds for each agent.  The former is important for ensuring that an appropriate number of agents are capable of responding to various levels of demand for each task while the latter regulates whether an agent must select between multiple tasks for which their thresholds are satisfied.

The experiments performed with shuffled thresholds demonstrate the relevance of the distribution of thresholds across tasks for an agent.  The shuffled data elements in Figures~\ref{fig:thresh-types} and ~\ref{fig:switches} demonstrate how changing the relative values of task thresholds for an agent affect swarm performance.  As previously described, these elements represent runs using evolved response thresholds that have been randomly permuted, maintaining the set of thresholds for each task but redistributing them among agents.  This disrupts the relative values of thresholds for an agent across the tasks.  Both domain problem performance and specialization are significantly decreased for shuffled thresholds.

The importance of threshold distribution for a task is evident in Figure ~\ref{fig:timelapse}.  The first and final histograms show the evolved change in distribution.  The initial values, uniformly distributed at random, have 18 thresholds less than $0.35$ and 8 less than $0.20$.  By generation 500, the evolved thresholds have 26 thresholds less than $0.35$ and 19 less than $0.20$.  With many more agents responding to lower demand, the tracker is able to maintain a much smaller distance to the target, by more than a factor of 10, as seen in the legend of each histogram.

%% file: sec-conclusion.tex
\section{Conclusion}
\label{sec:conclusion}

In this paper we examine evolved response thresholds for a dynamic task allocation problem in a decentralized, deterministic response-based swarm.  The testbed problem we use allows creation of task demands with significant variation in balance of demands between tasks, rate of change in demand, and magnitude of change in demand.  Our main findings are:
\begin{itemize}
\item With evolved response thresholds, swarm performance is significantly improved relative to randomly distributed thresholds.  The improvement is reflected in both domain goals and agent specialization.
\item Thresholds evolved for some problem instances, \simpath{circle} and \simpath{random}, generalize to other problem instances with very different task demands.
\item \simpath{circle} is a simple deterministic problem instance that generalizes to all other instances tested.  This is true independent of the radius used for training and with as little as one revolution used for fitness evaluation during evolution.
\item Generalization derives from variation in the balance of simultaneous task demand during evolution.
\item The distribution of thresholds for a task, across all agents, is an important factor in swarm success with respect to appropriately addressing task demands.
\item The distribution of thresholds for an agent, across the tasks, is an important factor in swarm success with respect to appropriately addressing task demands and for developing agent specialization.
\end{itemize}

One drawback of using evolved response thresholds is the expense of generating them.  The time required would likely make their use unrealistic if each problem instance required custom thresholds.  Our results provide hope that for the existence of universal training instances, problem instances that result in evolved thresholds that can be effective for a variety of other problem instances.

The most significant remaining question is to what degree these results hold for other problems.  The task allocation problem we use as a testbed is very general in its ability to model different task demands, however, testing on other problems is needed.  In future work, we will implement a very general dynamic task allocation problem and attempt to demonstrate the existence of universal training instances in that domain.

%% file: evolved-thresholds.bbl
\begin{thebibliography}{10}

\bibitem{ampatzis2008}
Christos Ampatzis, Elio Tuci, Vito Trianni, and Marco Dorigo.
\newblock Evolution of signaling in a multi-robot system: Categorization and
  communication.
\newblock {\em Adaptive Behavior}, 16:5--26, 2008.

\bibitem{ashby1958}
W.~Ross Ashby.
\newblock Requisite variety and its implications for the control of complex
  systems.
\newblock {\em Cybernetica}, 1(2):83--99, 1958.

\bibitem{aubert-kato2017}
Nathanael Aubert-Kato, Charles Fosseprez, Guillaume Gines, Ibuki Kawamata, Huy
  Dinh, Leo Cazenille, Andre Estevez-Tores, Masami Hagiya, Yannick Rondelez,
  and Nicolas Bredeche.
\newblock Evolutionary optimization of self-assembly in a swarm of
  bio-micro-robots.
\newblock In {\em Proc. of the Genetic and Evolutionary Computation
  Conference}, pages 59--66, 2017.

\bibitem{baldassarre2003}
Gianluca Baldassarre, Stefano Nolfi, and Domenico Parisi.
\newblock Evolving mobile robots able to display collective behaviors.
\newblock {\em Artificial Life}, 9:255--267, 2003.

\bibitem{baldassarre2007}
Gianluca Baldassarre, Vito Trianni, Michael Bonani, Francesco Mondada, Marco
  Dorigo, and Stefano Nolfi.
\newblock Self-organized coordinated motion in groups of physically connected
  robots.
\newblock {\em IEEE Transactions on Systems, Man, and Cybernetics -- Part B:
  Cybernetics}, 37:224--239, 2007.

\bibitem{beckman2008}
Benjamin~E. Beckman and Philip~K. McKinley.
\newblock Evolution of adaptive population control in multi-agent systems.
\newblock In {\em Proc. 2nd IEEE Intl. Conf. on Self-Adaptive and
  Self-Organizing Systems}, 2008.

\bibitem{beni1992}
Gerardo Beni.
\newblock Distributed robotic systems and swarm intelligence.
\newblock {\em J Robotic Soc Jpn}, 10:31--37, 1992.

\bibitem{bonabeau1996}
Eric Bonabeau, Guy Theraulaz, and Jean-Louis Deneubourg.
\newblock Quantitiate study of the fixed threshold model for the regulation of
  division of labor in insect societies.
\newblock {\em Proc. Royal Society of London: Biological Sciences},
  263(1376):1565--1569, 1996.

\bibitem{bonabeau1998}
Eric Bonabeau, Guy Theraulaz, and Jean-Louis Deneubourg.
\newblock Fixed response thresholds and the regulation of division of labor in
  insect societies.
\newblock {\em Bulletin of Mathematical Biology}, 60:753--807, 1998.

\bibitem{brutschy2012}
Arne Brutschy, Nam-Luc Tran, Nadir Baiboun, Marco Frison, Giovanni Pini, Andrea
  Roli, Marco Dorigo, and Mauro Birattari.
\newblock Costs and benefits of behavioral specialization.
\newblock {\em Robotics and Autonomous Systems}, 60:1408--1420, 2012.

\bibitem{campbell2011}
Adam Campbell, Cortney Riggs, and Annie~S. Wu.
\newblock On the impact of variation on self-organizing systems.
\newblock In {\em Proc. 5th IEEE Int'l Conf. Self-Adaptive and Self-Organizing
  Systems}, 2011.

\bibitem{campos2000}
Mike Campos, Eric Bonabeau, Guy Theraulaz, and {Jean-Louis} Deneubourg.
\newblock Dynamic scheduling and division of labor in social insects.
\newblock {\em Adaptive Behavior}, 8:83--96, 2000.

\bibitem{castello2018}
E.~Castello, T.~Yamamoto, F.~D. Libera, W.~Liu, A.~F.~T. Winfield, Y.~Nakamura,
  and H.~Ishiguro.
\newblock Adaptive foraging for simulated and real robotic swarms: The
  dynamical response threshold approach.
\newblock {\em Swarm Intelligence}, 10:1--31, 2018.

\bibitem{castello2013}
Eduardo Castello, Tomoyuki Yamamoto, Yutaka Nakamura, and Hiroshi Ishiguro.
\newblock Task allocation for a robotic swarm based on an adaptive response
  threshold model.
\newblock In {\em Proc. 13th Int'l Conf. Control, Automation, and Systems},
  pages 259--266, 2013.

\bibitem{cicirello2002}
V.~A. Cicirello and S.~F. Smith.
\newblock Distributed coordination of resources via wasp-like agents.
\newblock In {\em Workshop on Radical Agent Concepts, LNAI 2564}, pages 71--80,
  2002.

\bibitem{cicirello2003}
Vincent~A. Cicirello and Stephen~F. Smith.
\newblock Distributed coordination of resources via wasp-like agents.
\newblock In {\em Lecture Notes in Artificial Intelligence}, volume 2564, pages
  71--80, 2002.

\bibitem{correll2008}
N.~Correll.
\newblock Parameter estimation and optimal control of swarm-robotic systems: A
  case study in distributed task allocation.
\newblock In {\em Proceedings of the IEEE International Conference on Robotics
  and Automation}, pages 3302--3307, 2008.

\bibitem{delope2013}
Javier {de Lope}, Dario Maravall, and Yadira Quinonez.
\newblock Response threshold models and stochastic learning automata for
  self-coordination of heterogeneous multi-task distribution in multi-robot
  systems.
\newblock {\em Robotics and Autonomous Systems}, 61:714--720, 2013.

\bibitem{delope2015}
Javier {de Lope}, Dario Maravall, and Yadira Quinonez.
\newblock Self-organizing techniques to improve the decentralized multi-task
  distribution in multi-robot systems.
\newblock {\em Neurocomputing}, 163:47--55, 2015.

\bibitem{nsga-ii}
Kalyanmoy Deb, Amrit Pratap, Sameer Agarwal, and T.~Meyarivan.
\newblock A fast and elitist multiobjective genetic algorithm: Nsga-ii.
\newblock {\em IEEE Transactions on Evolutionary Computation}, 6(2):182--197,
  2002.

\bibitem{dorigo2004}
Marco Dorigo, Vito Trianni, Erol Sahin, Roderich Gro{\ss}, Thomas~H. Labella,
  Gianluca Baldassarre, Stefano Nolfi, Jean-Louis Deneubourg, Francesco
  Mondada, Dario Floreano, and Luca~M. Gambardella.
\newblock Evolving self-organizing behaviors for a swarm-bot.
\newblock {\em Autonomous Robots}, 17:223--245, 2004.

\bibitem{dossantos2009}
F.~dos Santos and A.~L.~C. Bazzan.
\newblock An ant based algorithm for task allocation in large-scale and dynamic
  multiagent scenarios.
\newblock In {\em Proceedings of the Genetic and Evolutionary Computation
  Conference}, pages 73--80, 2009.

\bibitem{duarte2012}
Ana Duarte, Ido Pen, Laurent Keller, and Franz~J. Weissing.
\newblock Evolution of self-organized division of labor in a response threshold
  model.
\newblock {\em Behavioral Ecology and Sociobiology}, 66:947--957, 2012.

\bibitem{duarte2014}
M.~Duarte, Sancho Oliveira, and A.~Christensen.
\newblock Hybrid control for large swarms of aquatic drones.
\newblock In {\em Proceedings of the Fourteenth Intl. Conf. on the Synthesis
  and Simulation of Living Systems}, pages 785--792, 2014.

\bibitem{duarte2016a}
Miguel Duarte, Vasco Costa, Jorge Gomes, Tiago Rodrigues, Fernando Silva,
  Sancho~Moura Olivieira, and Anders~Lyhne Christensen.
\newblock Evolution of collective behaviors for a real swarm of aquatic surface
  robots.
\newblock {\em PLoS ONE}, 11(3), 2016.

\bibitem{duarte2016b}
Miguel Duarte, Jorge Gomes, Vasco Costa, Sancho~Moura Oliveira, and
  Anders~Lyhne Christensen.
\newblock Hybrid control for a real swarm robotics system in an intruder
  detection task.
\newblock In {\em Proceedings of the European Conference on the Applications of
  Evolutionary Computation}, pages 213--230, 2016.

\bibitem{ferrante2015}
Eliseo Ferrante, Ali~Emre Turgut, Edgar~Du\'{e}\ {n}ez Guzm\'{a}n, Marco
  Dorigo, and Tom Wenseleers.
\newblock Evolution of self-organized task specialization in robot swarms.
\newblock {\em PLOS Computational Biology}, 11, 2015.

\bibitem{fischer2018}
Dominik Fischer, Sanaz Mastaghim, and Larissa Albantakis.
\newblock How swarm size during evolution imacts the behavior,
  generalizability, and brain connectivity of animats performing a spatial
  navigation task.
\newblock In {\em Proc. Genetic and Evolutionary Computation Conference}, pages
  77--84, 2018.

\bibitem{forrest1993}
Stephanie Forrest and Melanie Mitchell.
\newblock What makes a problem hard for a genetic algorithm? some anomalous
  results and their explanation.
\newblock {\em Machine Learning}, 13:285--319, 1993.

\bibitem{fortin2012}
F{\'e}lix-Antoine Fortin, Fran{\c{c}}ois-Michel~De Rainville, Marc-Andr{\'e}
  Gardner, Marc Parizeau, and Christian Gagn{\'e}.
\newblock Deap: Evolutionary algorithms made easy.
\newblock {\em Journal of Machine Learning Research}, 13(Jul):2171--2175, 2012.

\bibitem{gautrais2002}
Jacques Gautrais, Guy Theraulaz, {Jean-Louis} Deneubourg, and Carl Anderson.
\newblock Emergent polyethism as a consequence of increase colony size in
  insect societies.
\newblock {\em Journal of Theoretical Biology}, 215:363--373, 2002.

\bibitem{goldingay2013}
H.~Goldingay and J.~van Mourik.
\newblock The effect of load on agent-based algorithms for distributed task
  allocation.
\newblock {\em Information Sciences}, 222:66--80, 2013.

\bibitem{goldsby2012}
Heather~J. Goldsby, Anna Dornhaus, Benjamin Kerr, and Charles Ofria.
\newblock Task-switching costs promote the evolution of division of labor and
  shifts in individuality.
\newblock {\em Proceedings of the National Academy of Sciences},
  109(34):13686--13691, 2012.

\bibitem{goldsby2010}
Heather~J. Goldsby, David~B. Knoester, and Charles Ofria.
\newblock Evolution of division of labor in genetically homogeneous groups.
\newblock In {\em Proceedings of the 2010 Genetic and Evolutionary Computation
  Conference (GECCO)}, 2010.

\bibitem{gomes2013b}
Jorge Gomes and Anders~L. Christensen.
\newblock Generic behaviour similarity measures for evolutionary swarm
  robotics.
\newblock In {\em Proceedings of Genetic and Evolutionary Computation
  Conference (GECCO)}, pages 199--206, 2013.

\bibitem{gomes2013a}
Jorge Gomes, Paulo Urbano, and Anders~Lyhne Christensen.
\newblock Evolution of swarm robotics systems with novelty search.
\newblock {\em Swarm Intelligence}, 7:115--144, 2013.

\bibitem{gross2008}
Roderich Gro\ss{} and Marco Dorigo.
\newblock Evolution of solitary and group transport behaviors for autonomous
  robots capable of self-assembling.
\newblock {\em Adaptive Behavior}, 16:285--305, 2008.

\bibitem{guo2020}
Miao Guo, Bin Xie, Jie Chen, and Yipeng Wang.
\newblock Multi-agent coalition formation by an efficient genetic algorithm
  with heuristic initialization and repair strategy.
\newblock {\em Swarm and Evolutionary Computation}, 55, 2020.

\bibitem{hart2018}
Emma Hart, Andreas S.~W. Steyven, and Ben Paechter.
\newblock Evolution of a functionally diverse swarm via a novel decentralised
  quality-diversity algorithm.
\newblock In {\em Proc. Genetic and Evolutionary Computation Conference}, pages
  101--108, 2018.

\bibitem{hauert2009}
Sabine Hauert, Jean-Christophe Zuffery, and Dario Floreano.
\newblock Evolved swarming without positioning information: an application in
  aerial communication relay.
\newblock {\em Autonomous Robots}, 26:21--32, 2009.

\bibitem{holbrook2011}
C.~Tate Holbrook, Phillip~M. Barder, and Jennifer~H. Fewell.
\newblock Division of labor increases with colony size in the harvester ant
  pogonomyrmex californicus.
\newblock {\em Behavioral Ecology}, 22:960--966, 2011.

\bibitem{holbrook2013}
C.T. Holbrook, T.H. Eriksson, R.P. Overson, J.~Gadau, and J.H. Fewell.
\newblock Colony-size effects on task organization in the harvester ant
  pogonomymex californicus.
\newblock {\em Insectes Sociaux}, 60(2):191--201, 2013.

\bibitem{huang2017}
Chien-Lun Hunag and Geoff Nitschke.
\newblock Evolving collective driving behaviors.
\newblock In {\em Proc. 16th Intl. Conf. Autonomous Agents and MultiAgent
  Systems}, pages 1573--1574, 2017.

\bibitem{jeanne1986}
Robert~L. Jeanne.
\newblock The evolution of the organization of work in social insects.
\newblock {\em Monitore Zoologico Italiano}, 20:119--133, 1986.

\bibitem{jeanson2014}
Rapha\"el Jeanson and Anja Weidenm\"uller.
\newblock Interindividual variability in social insects - proximate causes and
  ultimate consequences.
\newblock {\em Biological Reviews}, 89:671--687, 2014.

\bibitem{kalra2006}
Nidhi Kalra and Alcherio Martinoli.
\newblock A comparative study of market-based and threshold-based task
  allocation.
\newblock In {\em Distributed Autonomous Robotics Systems 7}, pages 91--101,
  2006.

\bibitem{kanakia2016}
Anshul Kanakia, Behrouz Touri, and Nikolaus Correll.
\newblock Modeling multi-robot task allocation with limited information as
  global game.
\newblock {\em Swarm Intelligence}, 10:147--160, 2016.

\bibitem{kazakova2018}
Vera~A. Kazakova and Annie~S. Wu.
\newblock Specialization vs. re-specialization: Effects of {Hebbian} learning
  in a dynamic environment.
\newblock In {\em Proc 31st FLAIRS}, pages 354--359, 2018.

\bibitem{kazakova2020}
Vera~A. Kazakova, Annie~S. Wu, and Gita~R. Sukthankar.
\newblock Respecializing swarms by forgetting reinforced thresholds.
\newblock {\em Swarm Intelligence}, 2020.

\bibitem{kittithreerapronchai2003}
Oran Kittithreerapronchai and Carl Anderson.
\newblock Do ants paint trucks better than chickens? {Market} versus response
  threshold for distributed dynamic scheduling.
\newblock In {\em Proceedings of the Congress on Evolutionary Computation},
  pages 1431--1439, 2003.

\bibitem{krieger2000a}
Michael J.~B. Krieger, Jean-Bernard Billeter, and Laurent Keller.
\newblock Ant-like task allocation and recruitment in cooperative robots.
\newblock {\em Nature}, 406:992--995, 2000.

\bibitem{krieger2000b}
Michael J.~B. Krieger, Jean-Bernard Billeter, and Laurent Keller.
\newblock Ant-like task allocation and recruitment in cooperative robots.
\newblock {\em Nature}, 406:992--995, 2000.

\bibitem{labella2006}
Thomas~H. Labella, Marco Dorigo, and {Jean-Louis} Deneubourg.
\newblock Division of labor in a group of robotics inspired by ants' foraging
  behavior.
\newblock {\em {ACM} Transactions on Autonomous and Adaptive Systems},
  1(1):4--25, 2006.

\bibitem{langridge2004}
Elizabeth~A. Langridge, Nigel~R. Franks, and Ana~B. Sendova-Franks.
\newblock Improvement in collective performance with experience in ants.
\newblock {\em Behavioral Ecology and Sociobiology}, 56:523--529, 2004.

\bibitem{merkle2004}
Daniel Merkle and Martin Middendorf.
\newblock Dynamic polyethism and competition for tasks in threshold
  reinforcement models of social insects.
\newblock {\em Adaptive Behavior}, 12(3-4):251--262, 2004.

\bibitem{meyer2015}
Bernd Meyer, Anja Weidenmuller, Rui Chen, and Julian Garcia.
\newblock Collective homeostasis and time-resolved models of self-organised
  task allocation.
\newblock In {\em Proc. 9th EAI Int'l Conf Bio-inspired Info \& Comm Tech},
  pages 469--478, 2015.

\bibitem{moritz2015}
Ruby~L. Moritz and Martin Middendorf.
\newblock Evolutionary inheritance mechanisms for multi-criteria decision
  making in multi-agent systems.
\newblock In {\em Proc. Genetic and Evolutionary Computation Conference}, pages
  65--72, 2015.

\bibitem{narzisi2006}
Giuseppe Narzisi, Venkatesh Mysore, and Bud Mishra.
\newblock Multi-objective evolutionary optimization of agent-based models: An
  application to emergency response planning.
\newblock In {\em 2nd Intl. Conf. on Computational Intelligence}, pages
  224--230, 2006.

\bibitem{neupane2019}
Aadesh Neupane and Michael~A. Goodrich.
\newblock Designing emergent swarm behaviors usign behavior trees and
  grammatical evolution.
\newblock In {\em Proc. 18th Intl. Conf. Autonomous Agents and MultiAgent
  Systems}, pages 2138--2140, 2019.

\bibitem{neupane2018}
Aadesh Neupane, Michael~A. Goodrich, and Eric~G. Mercer.
\newblock Geese: Gramatical evolution algorithm for evolution of swarm
  behaviors.
\newblock In {\em Proc. Genetic and Evolutionary Computation Conference}, pages
  999--1006, 2018.

\bibitem{niccolini2010}
M.~Niccolini, M.~Innocenti, and L.~Pollini.
\newblock Multiple uav task assignment using descriptor functions.
\newblock In {\em Proceedings of the 18th IFAC Symposium on Automatic Control
  in Aerospace}, pages 93--98, 2010.

\bibitem{nitschke2009}
G.~Nitschke.
\newblock Neuro-evolution methods for gathering and collective construction.
\newblock In {\em Proceedings of the 10th European Conference on Artificial
  Life}, pages 111--119, 2009.

\bibitem{nitschke2012b}
G.S. Nitschke, A.E. Eiben, and M.C. Schut.
\newblock Evolving team behaviors with specialization.
\newblock {\em Genetic Programming and Evolvable Machines}, 13:493--536, 2012.

\bibitem{nitschke2012a}
G.S. Nitschke, M.C. Schut, and A.E. Eiben.
\newblock Evolving behavioral specialization in robot teams to solve a
  collective construction task.
\newblock {\em Swarm and Evolutionary Computation}, 2:25--38, 2012.

\bibitem{nouyan2005}
Shervin Nouyan, Roberto Ghizzioli, Mauro Birattari, and Marco Dorigo.
\newblock An insect-based algorithm for the dynamic task allocation problem.
\newblock Technical report, IRIDIA, 2005.
\newblock TR/IRIDIA/2005-031.

\bibitem{panait2006}
Liviu Panait, Sean Luke, and R.~Paul Wiegand.
\newblock Biasing coevolutionary search for optimal multiagent behaviors.
\newblock {\em IEEE Transactions on Evolutionary Computation}, 10(6):629--645,
  2006.

\bibitem{pang2017}
B.~Pang, C.~Zhang, Y.~Song, and H.~Wang.
\newblock Seld-organized task allocation in swarm robotics foraging based on
  dynamical response threshold approach.
\newblock In {\em Proceedings of the 18th International Conference on Advanced
  Robotics}, pages 256--261, 2017.

\bibitem{pini2008}
Giovanni Pini and Elio Tuci.
\newblock On the design of neuro-controllers for individual and social learning
  behaviour in autonomous robots: an evolutionary approach.
\newblock {\em Connection Science}, 20:211--230, 2008.

\bibitem{price2004}
Richard Price and Peter Tino.
\newblock Evaluation of adaptive nature inspired task allocation against
  alternative decentralised multiagent strategies.
\newblock In {\em Proceedings of the Parallel Problem Solving from Nature, LNCS
  3242}, pages 982--990, 2004.

\bibitem{ravary2007}
Fabien Ravary, Emmanuel Lecoutey, Gwenael Kaminski, Nicolas Chaline, and Pierre
  Jaisson.
\newblock Individual experience alone can generate lasting division of labor in
  ants.
\newblock {\em Current Biology}, 17:1308--1312, 2007.

\bibitem{riggs2012}
Cortney Riggs and Annie~S. Wu.
\newblock Variation as an element in multi-agent control for target tracking.
\newblock In {\em Proc. IEEE/RSJ Int'l Conf. Intelligent Robots and Systems},
  pages 834--841, 2012.

\bibitem{samarasinghe2018}
Dilini Samarasinghe, Erandi Lakshika, Michael Barlow, and Kathryn Kasmarik.
\newblock Automatic synthesis of swarm behavioural rules from their atomic
  components.
\newblock In {\em Proc. Genetic and Evolutionary Computation Conference}, pages
  133--140, 2018.

\bibitem{soysal2007}
Onur Soysal, Erkin Bah\c{c}ec\.{i}, and Erol \c{S}ah\.{i}n.
\newblock Aggregation in swarm robotic systems: Evolution and probabilistic
  control.
\newblock {\em Turkish Journal of Electrical Engineering and Computer
  Sciences}, 15:199--225, 2007.

\bibitem{sperati2008}
Valerio Sperati, Vito Trianni, and Stefano Nolfi.
\newblock Evolving coordinated group behaviours through maximisation of mean
  mutual information.
\newblock {\em Swarm Intelligence}, 2:73--95, 2008.

\bibitem{sperati2011}
Valerio Sperati, Vito Trianni, and Stefano Nolfi.
\newblock Self-organised path formation in a swarm of robots.
\newblock {\em Swarm Intelligence}, 5:97--119, 2011.

\bibitem{steyven2017}
Andreas Steyven, Emma Hart, and Ben Paechter.
\newblock An investigation of environmental influence on the benefits of
  adaptation mechanisms in evolutionary swarm robotics.
\newblock In {\em Proc. Genetic and Evolutionary Computation Conference}, pages
  155--162, 2017.

\bibitem{theraulaz1998}
Guy Theraulaz, Eric Bonabeau, and {Jean-Louis} Deneubourg.
\newblock Response threshold reinforcement and division of labour in insect
  societies.
\newblock {\em Proc. Royal Society B}, 265:327--332, 1998.

\bibitem{theraulaz1991}
Guy Theraulaz, Simon Goss, Jacques Gervet, and Jean-Louis Deneubourg.
\newblock Task differentiation in polistes wasp colonies: A model for
  self-organizing groups of robots.
\newblock In {\em Proceedings of the 1st International Conference on Simulation
  of Adaptive Behavior: From Animals to Animats}, pages 346--355, 1991.

\bibitem{trianni2003}
Vito Trianni, Roderich Gro\ss{}, Thomas~H. Labella, Erol \c{S}ah\.{i}n, and
  Marco Dorigo.
\newblock Evolving aggregation behaviors in a swarm of robots.
\newblock In {\em Proceedings of the European Conference on Artificial Life},
  pages 865--874, 2003.

\bibitem{trianni2006}
Vito Trianni, Stefano Nolfi, and Marco Dorigo.
\newblock Cooperative hole avoidance in a \textit{swarm-bot}.
\newblock {\em Robotics and Autonomous Systems}, 54:97--103, 2006.

\bibitem{tuci2014}
Elio Tuci.
\newblock Evolutionary swarm robotics: Genetic diversity, task allocation and
  task switching.
\newblock In {\em Proceedings of the 9th International Conference on Swarm
  Intelligence (ANTS)}, pages 148--160, 2014.

\bibitem{wang2019}
Jane~X. Wang, Edward Hughes, Christantha Fernando, Wojciech~M. Czaarnecki,
  Edgar~A. Duenez-Guzman, and Joel~Z. Leibo.
\newblock Evolving intrinsic motivations for altruistic behavior.
\newblock In {\em Proc. 18th Intl. Conf. Autonomous Agents and MultiAgent
  Systems}, pages 683--692, 2019.

\bibitem{weidenmuller2004}
Anja {Weidenm\"uller}.
\newblock The control of nest climate in bumblebee (${B}ombus$ $terrestris$)
  colonies: Interindividual variability and self reinforcement in fanning
  response.
\newblock {\em Behavioral Ecology}, 15:120--128, 2004.

\bibitem{weidenmuller2019}
Anja {Weidenm\"uller}, Rui Chen, and Bernd Meyer.
\newblock Reconsidering response threshold models -- short-term response
  patterns in thermoregulating bumblebees.
\newblock {\em Behavioral Ecology and Sociobiology}, 73, 2019.

\bibitem{wu2020ants}
Annie~S. Wu and H.~David Mathias.
\newblock Dynamic response thresholds: Heterogeneous ranges allow
  specialization while mitigating convergence to sink states.
\newblock In {\em Proceedings of the 12th International Conference on Swarm
  Intelligence}, pages 107--120, 2020.

\bibitem{wu2020a}
Annie~S. Wu, H.~David Mathias, Joseph~P. Giordano, and Anthony Hevia.
\newblock Effects of response threshold distribution on dynamic division of
  labor in decentralized swarms.
\newblock In {\em Proc. 33rd Int'l Florida Artificial Intelligence Research
  Society Conference}, 2020.

\bibitem{wu2021}
Annie~S. Wu, H.~David Mathias, Joseph~P. Giordano, and Arjun Pherwani.
\newblock Collective control as a decentralized task allocation testbed.
\newblock Technical Report CS-TR-21-01, University of Central Florida, 2021.

\bibitem{wu2018}
Annie~S. Wu and Cortney Riggs.
\newblock Inter-agent variation improves dynamic decentralized task allocation.
\newblock In {\em Proc. 31st Int'l Florida Artificial Intelligence Research
  Society Conference}, pages 366--369, 2018.

\bibitem{yamada2013}
Naoki Yamada and Chiaka Sakama.
\newblock Evolution of self-interested agents: An experimental study.
\newblock In {\em 7th Intl. Workshop Multi-disciplinary Trends in AI}, pages
  329--340, 2013.

\bibitem{yang2010}
Yongming Yang, Xihui Chen, and Qingjun Li.
\newblock Swarm robots task allocation based on local communication.
\newblock In {\em Proceedings of the International Conference on Computer,
  Mechatronics, Control, and Electronic Engineering}, pages 415--418, 2010.

\bibitem{yu2011}
Ling Yu, Zhiqi Shen, Chunyan Miao, and Victor Lesser.
\newblock Genetic algorithm aided optimization of hierarchical multiagent
  system organization.
\newblock In {\em Proc. 10th Intl. Conf. Autonomous Agents and MultiAgent
  Systems}, pages 1169--1170, 2011.

\end{thebibliography}
